\documentclass[journal]{IEEEtran}
\IEEEoverridecommandlockouts
\usepackage{cite}
\usepackage{amsmath,amssymb,amsfonts}
\usepackage{balance}
\usepackage{algorithmic}
\usepackage{graphicx}
\usepackage{textcomp}
\usepackage{makeidx} 
\usepackage{mdwtab}
\usepackage{algorithm}
\usepackage{algorithmic}
\usepackage{xcolor}
\usepackage{amsmath}
\usepackage{tabularx}
\usepackage{dirtree}
\usepackage{tikz}
\usepackage{xcolor}
\usepackage[flushleft]{threeparttable}
\usepackage{graphicx}
\graphicspath{{./images/}}

\usepackage{textcomp}
\usepackage{tabularx}
\usepackage{multirow}
\usepackage{multicol}
\usepackage{dcolumn}
\usepackage{booktabs}
\usepackage{comment}
\usepackage{soul}
\usepackage{ulem}
\usepackage{enumerate}

\usepackage{float}

\usepackage{caption}
\usepackage{subcaption}

\def\BibTeX{{\rm B\kern-.05em{\sc i\kern-.025em b}\kern-.08em
    T\kern-.1667em\lower.7ex\hbox{E}\kern-.125emX}}
\begin{document}
 
\title{Towards 6G Internet of Things: Recent Advances, Use Cases, and Open Challenges}
 
\author{\vspace{4 mm}Zakria~Qadir, Hafiz~Suliman~Munawar, Nasir Saeed, ~\IEEEmembership{Senior~Member,~IEEE} and Khoa Le,~\IEEEmembership{Senior~Member,~IEEE} 
         
\IEEEcompsocitemizethanks{\IEEEcompsocthanksitem Zakria Qadir and Khoa Le is with the School of Engineering, Design and Built Environment, Western Sydney University, Locked Bag 1797, Penrith, NSW 2751, Australia.\IEEEcompsocthanksitem Hafiz Suliman Munawar is a PhD student at the University of New South Wales, Kensington, Sydney, NSW 2052, Australia.
\IEEEcompsocthanksitem Nasir Saeed is with the Department of Electrical Engineering, National University of Technology, Islamabad, Pakistan.
\\ 
%e-mail: \{Z.Qadir@westernsydney.edu.au}% <-this % stops a space
}}

\maketitle

\begin{abstract}
Smart services based on the Internet of Everything (IoE) are gaining considerable popularity due to the ever-increasing demands of wireless networks. This demands the appraisal of the wireless networks with enhanced properties as next-generation communication systems. Although 5G networks show great potential to support numerous IoE based services, it is not adequate to meet the complete requirements of the new smart applications. Therefore, there is an increased demand for envisioning the 6G wireless communication systems to overcome the major limitations in the existing 5G networks. 
Moreover, incorporating artificial intelligence in 6G will provide solutions for very complex problems relevant to network optimization. Furthermore, to add further value to the future 6G networks, researchers are investigating new technologies, such as THz and quantum communications. The requirements of future 6G wireless communications demand to support massive data-driven applications and the increasing number of users. This paper presents recent advances in the 6G wireless networks, including the evolution from 1G to 5G communications, the research trends for 6G, enabling technologies, and state-of-the-art 6G projects. 
\end{abstract}

 
\section{Introduction}
The up-gradation of mobile communication systems to a more advanced generation usually occurs with every turn of decade \cite{chen2020vision}. Following the usual convention, in 2020, mobile communication systems entered into fifth-generation (5G) since its inception in the 1980s. 5G is dubbed by many as the pinnacle of mobile communication technology  \cite{aka}. 5G and its preceding fourth generation (4G, often known as LTE-Advanced) is known to build an Internet-of-Things (IoT) enabled intelligent services, and application-oriented eco-system \cite{kyi}. 
 \begin{table}
\caption{Description of the Symbols used in this article.}
\label{tab:description}
\begin{tabular}{p{1.5cm} p{6cm}} \toprule
{\textbf{Symbols}} & {\textbf{Description}} \\ \midrule
2D & Two Dimensional \\
3D & Three dimensional \\
5G & Fifth Generation \\
6G & Sixth Generation \\
%AF & amplify-and-forward \\
BER & Bit error rate \\
BS & Base Station \\
%CI & ? \\
CR & Cognitive radio \\
CoMP & Coordinate multipoint \\
D2D & Device-to-device \\
%DF & decode-and-forward \\
FDMA & Frequency Division Multiple Access \\
%FP & Fortress-Problem \\
EE & Energy efficiency\\
EH  & Energy harvesting\\
GD & Gradient descent \\
IPOPT & Interior point optimizer  \\
IoT & Internet of Things \\
ISR & Interference-to-signal ratio \\
LOS & Line of Sight \\
LTE-U & long-term evolution-unlicensed  \\
LIS & Large Intelligent Surfaces\\
IRS & Intelligent Reflecting Surfaces\\
%M2M & Machine-to-machine \\
MIMO & Multiple-input-multiple-output \\
MINLP & Mixed-integer non-linear programming  \\
MISO & Multiple-input single-output \\
MEC & Mobile edge computing\\
mmWave & Millimeter wave \\
NOMA & Non-orthogonal multiple access \\
OFDM & Orthogonal frequency division multiplexing \\
PSO & Particle swarm optimization \\
PU & Primary users \\
HCN & Heterogeneous cellular network\\
Het-IoT & Heterogeneous Internet of things\\
HVN & Heterogeneous vehicular network\\
QoE & Quality of experience\\
QoS & Quality of Service \\
RF & Radio frequency \\
RFIDs & Radio Frequency Identifications\\
RA & Resource allocation\\
RF & Radio frequency\\
SINR & Signal-to-interference-plus-noise ratio \\
SNR & Signal-to-noise ratio\\
SE & Spectrum efficiency\\
TDMA & Time Division Multiple Access \\
UAV & Unmanned aerial vehicles \\
V2V & Vehicle-to-vehicle\\
WLAN &  Wireless local area network\\
ZF & Zero-forcing\\
\end{tabular}
\end{table} 
More prominently, 5G offers a triad of characteristics, namely, enhanced Mobile BroadBand (eMBB), massive Machine Type Communications (mMTC), ultra-Reliable Low Latency Communications (uRLLC) that were particularly aimed to overcome the limitations of the 4G \cite{9380673}. For example, in comparison to 4G, 5G networks are expected to provide a peak data rate of 20 Gbps, 3x spectral efficiency, 100 times improved energy efficiency, and a Gbps user experience with an end-to-end latency of 1 ms \cite{9530717}. 5G would also support seamless connectivity for devices with mobility of 500 km/h, a connection density of 1 million devices/km$^2$, and an area traffic capacity of 10 Mbps/m$^2$ \cite{wwc}. The 5G networks have been anticipated to facilitate an extensive range of smart IoE related services; however, it will not be sufficient to meet the requirements of future smart communities \cite{ywh}.   
 
 \begin{figure*}[t]
\center
\includegraphics[width=\linewidth]{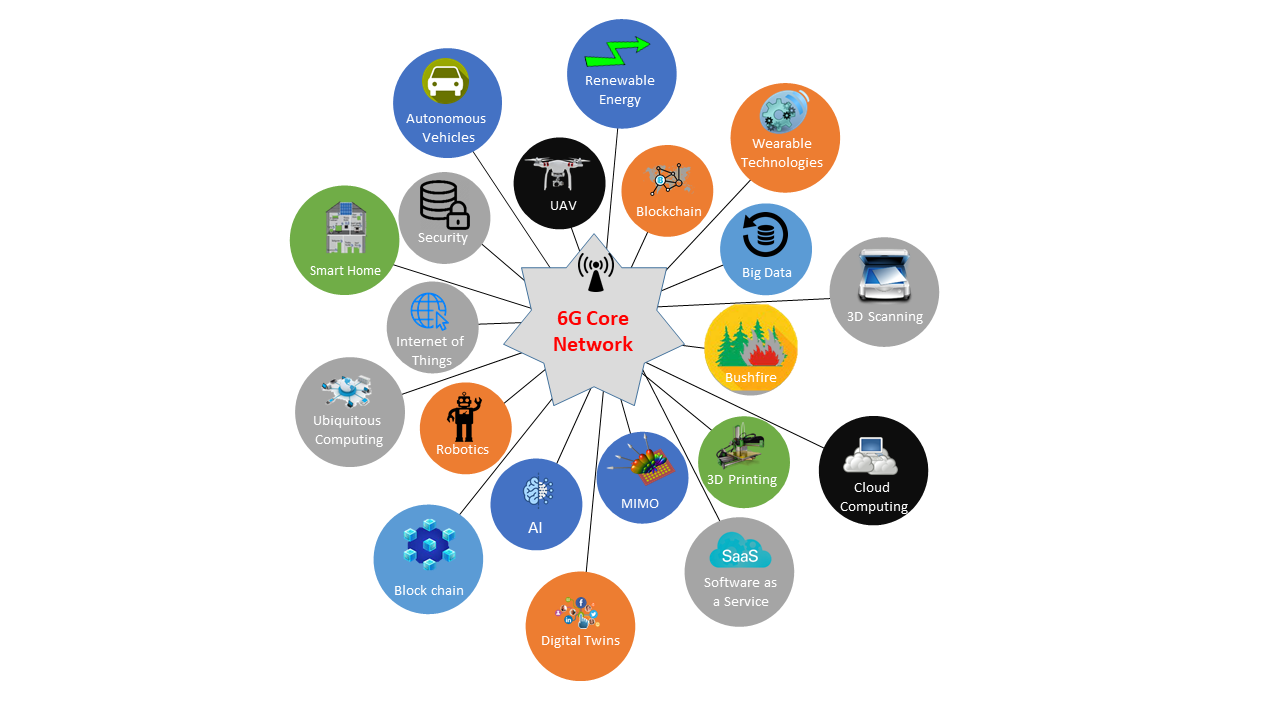} 
% \vspace{2.5 mm}
\caption{Envisioned 6G based applications.}
\label{Figure1}
% \vspace{-4 mm}
\end{figure*}

 \begin{table*}
\caption{Comparison of existing surveys.}
\label{table1}
%\begin{tabular}{p{1.5cm} p{1.05cm} p{1.5cm} p{1.3cm} p{1.0cm} p{1.55cm} p{1.75cm} p{1.35cm} p{1.35cm} p{1.65cm}}
\begin{tabular}{c|c|c|c|c|c|c|c|c|c|c|c|c|c|c|c|c|c|c}

\toprule
              R & IoT & AI & WI & ET & MEC & IS & DNS & HB & BA & BS & PC & CF & UAV & TC & OWC & MIMO & Sec. & BC \\ \midrule
\cite{9524814}   &     &  \checkmark  &    &    &     &    & \checkmark &    &    &    &    &    &     & \checkmark &   & \checkmark & \checkmark  &\checkmark \\
\cite{9509294}   &  \checkmark  &    &    &    &     & \checkmark &   &    &    &    &    &  \checkmark &  \checkmark   &    &       &      &   & \checkmark   \\
\cite{9358097}   &     &   & \checkmark &  \checkmark &     &    & \checkmark &    &    &    &  \checkmark  &    &     &    &      &      &   &   \\
\cite{9397776}   &  \checkmark  &    &    &    &     &    &   & \checkmark &    &    &    &    & \checkmark  &    &       &      &   &    \\
\cite{9403380}   &     & \checkmark  &    &    &     & \checkmark  & \checkmark &    &    &    &    &    &     & \checkmark & \checkmark   &      &   &   \\
\cite {9369324}  &  \checkmark & \checkmark  &    & \checkmark&   \checkmark & \checkmark &   &    &    &    &    &    &     & \checkmark   &      &      &   & \checkmark \\
\cite{9385374}   &  \checkmark &    &    & \checkmark & \checkmark   &\checkmark  &   &    &    &    &    & \checkmark &     &    &     &      &   &   \\
\cite{9184022}   & \checkmark & \checkmark  &  \checkmark &    &     & \checkmark &   &    &    &    &    &    &\checkmark  & \checkmark &     & \checkmark &\checkmark & \checkmark \\
\cite{9200376}   &  \checkmark &    & \checkmark &    &     &    & \checkmark & \checkmark& \checkmark  &    & \checkmark  &    &     & \checkmark      &   &      & \checkmark\\
\cite{9355403}   & \checkmark &\checkmark & \checkmark & \checkmark  &   \checkmark & \checkmark  &   &    &    &    &    &    & \checkmark &    & \checkmark &  \checkmark  & \checkmark& \checkmark\\
Our study        &  \checkmark & \checkmark  & \checkmark & \checkmark  &  \checkmark   &  \checkmark & \checkmark &  \checkmark  &\checkmark & \checkmark & \checkmark &  \checkmark & \checkmark  & \checkmark  & \checkmark  &  \checkmark   & \checkmark& \checkmark\\ \bottomrule
\end{tabular}
 
   \begin{tablenotes}
      \small
      \item {${}^\star$ R: Reference, IoT: Internet of Things, AI: Artificial Intelligence, WI: Wireless Information, ET: Energy Transfer, MEC: Mobile Edge Computing, IS: Integration of Sensing and Communication, DNS: Dynamic Network Slicing, HB: Holographic Beamforming, BA: Big Data Analytics, BS: Backscatter Communication, PC: Proactive Caching, CF: Cell-Free Communications, UAV: Unmanned Aerial Vehicle, TC: Terahertz Communications, OWC : Optical Wireless Communication, MIMO: Multiple Input Multiple Output, Sec.: Security, BC: Blockchain}.
    \end{tablenotes}
  \label{tab:coverage}%
\end{table*}%

 Since smart cities are automating our surroundings by enabling the digital layer on top of the traditional infrastructure,  the stakeholders' demand is abruptly increasing.   Therefore,   appropriate management for this digitization providing the ubiquitous solution for smart cities, disaster management, and other services is becoming critically important. Considering the forthcoming development in the domain of wireless technologies, particularly in smart cities, 5G may lack to meet future expectations as of 6G for the following facts:
\begin{itemize}
    \item as per the rapid growth of IoT devices in providing wireless connectivity to smart cities, there is an abrupt need of improvement to provide reliable connectivity to dense networks \cite{9328851}.
    \item introduction of flying cars, extended reality (XR), and telemedicine require high data transfer rate, low latency, and robustness for cellular networks that can only be possible with the envision 6G networks as shown in Figure \ref{Figure1}.
    \item it is believed that the future cellular networks will be robust, highly dynamic, complex, and embedded on ultra-large-scale chips. However, the current network architecture for both 4G and 5G is fixed to tackle a dedicated task only \cite{9044345}.  A state-of-the-art dynamic architecture is required in 6G that can  optimize based on the user demands.
\end{itemize}
 Table \ref{tab:description} shows description of the symbols used in this article.  
 
 \subsection{Related Surveys}
Many studies have focused on the 6G networks, facilitating technologies, architectures, and open research challenges in recent years. For instance, in \cite{9524814}, the authors portray a systematic review for 6G wireless communication based on the security and privacy perspective using blockchain technology. They have developed critical thinking for the architectural failure of a security system. Then, the authors in  \cite{9509294} discussed in depth the role of 6G communication for several IoT applications in the domain of healthcare, industries, autonomous vehicles, and satellite linkage using UAVs. 
Reference \cite{9358097} discusses main system model parameters like latency, energy consumption, network mobility.  The limitations of existing 5G communications are highlighted with the advancement of 6G communication in \cite{9397776}. 

In \cite{9403380}, the end-to-end transmission flow is surveyed with a focus on network access and robust routing control. Several machine learning applications are introduced with 6G aided IoT domain and blockchain for privacy and security perspective \cite{9369324}.  In \cite{9385374}, associated challenges related to terrestrial satellite networks are studied to overcome the performance parameters like channel fading, transmission delay, trajectory, and area coverage. 
Moreover, the use cases related to the 6G architecture and requirements are broadly categorized in \cite{9184022}. In \cite{9355403}, authors studied the resource allocation problems for next-generation heterogeneous networks considering the prospect of 6G.

\subsection{Main Contributions}
Unlike existing works, this survey addresses the state-of-the-art 6G wireless communication, recent advances, use cases, and open challenges. A detailed comparison between existing articles and with our survey is shown in Table \ref{table1}.  
 We focus on several important aspects of the envisioned 6G networks, such as robust connectivity, communication latency, edge computing, UAV application, and security issues. The collected literature review is from the past five years, focusing on the recent trends and future research directions. The main contribution of this survey are summarized as follows:
\begin{itemize}
    \item our main focus is to discuss in detail the important parameters of 6G technologies that were not fully optimized in 5G technology. This includes higher data rate, lower latency, improved reliability and accuracy, much higher energy efficiency, AI-IoT-based wireless connectivity, and 3D MIMO-oriented signal coverage. 
    \item the role of 6G in security and privacy is also studied particularly in the perspective of wireless connectivity.
    \item a systematic framework is designed to emphasize the applications of 6G in the domain of the smart home, smart industries, smart fire detection, smart parking, thus anticipating smart city concept. 
    \item an extensive comparison between 6G and the previous communication technologies is carried out to highlight the shortcomings in the previous architectures. 
\end{itemize}

\subsection{Organization of this paper}
The remaining paper is organized as follows: Section I discusses about the comparison between existing studies and why this survey would play a significant role for researchers in the context of 6G. Section II elaborates the evolution of mobile communication network from 1G to 6G. Section III extensively studies the research and marketing trend in the perspective of mobile communication network. Section IV highlights the network requirements for 6G communication network. Moreover, the essential enabling technologies for the 6G network are elaborated in the Section V. Finally, we present conclusion in Section VI. 

%{\color{red} \rule{\linewidth}{0.5mm} }

 \section{Evolution of  Mobile Communication Networks}
\begin{figure}
\center
\includegraphics[width=3in]{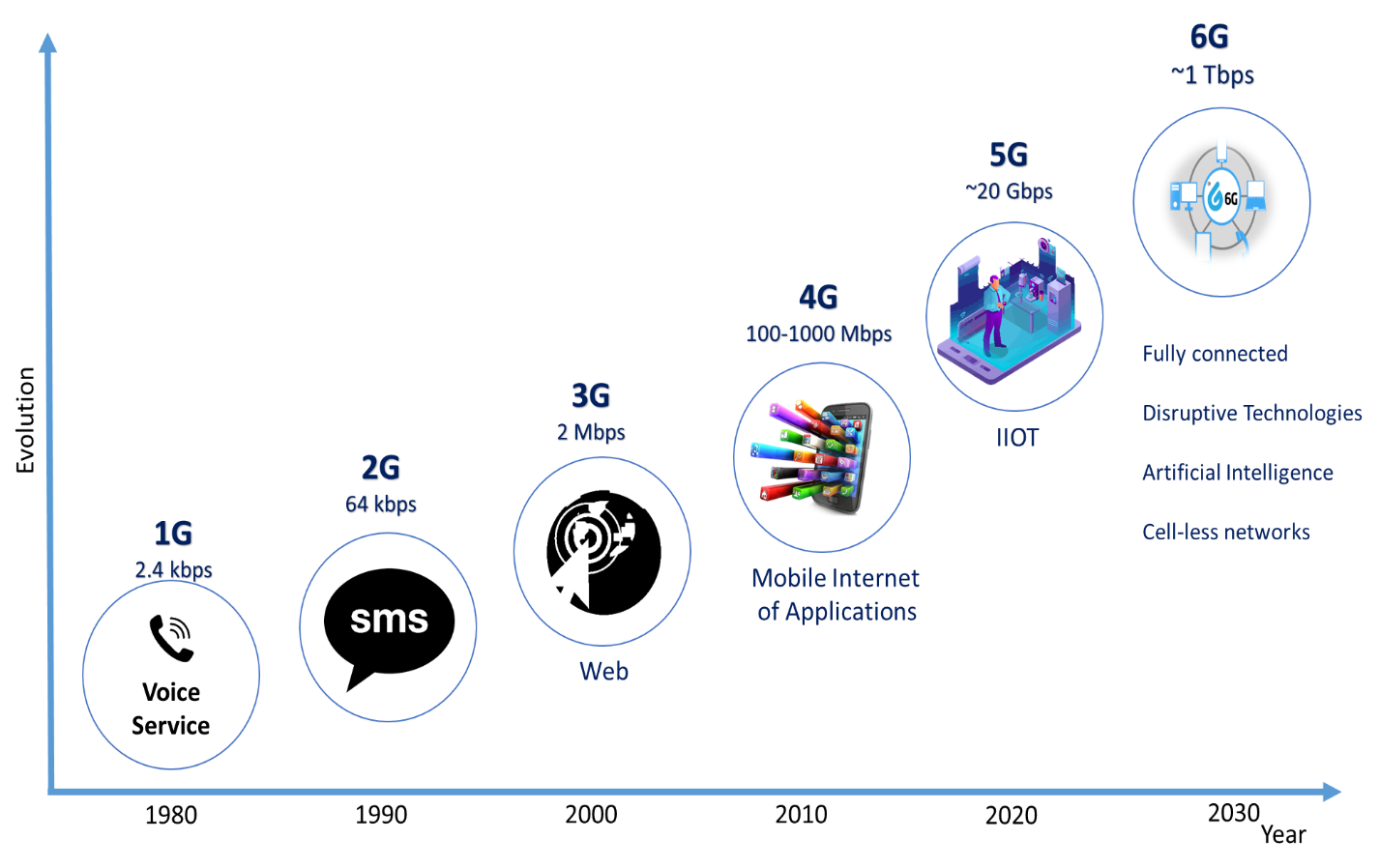} 
\vspace{2.5 mm}
\caption{Existing and expected mobile wireless communication evolution till 2030 }%\cite{hyw}}
\label{Figure 2}
\vspace{-4 mm}
\end{figure}

The recent advancements have led the mobile communication networks to bloom since the first development of the analog communications network during the 1980s. The considerable advances achieved in the mobile communication networks are not due to a one-step procedure but rather have been achieved through gradual changes over several generations in terms of the different aims, standards, capacities, perspectives, and technology applied to each existing generation. It is clearly established that new generations are introduced nearly ten years after the existing technologies \cite{hzl}. The evolution of mobile communication networks from 1G to 6G is depicted in Figure \ref{Figure 2} and the overall detailed comparison of the technologies is presented in Table \ref{Table 3}.

\subsection{Cellular Evolution: 1G TO 5G}
During the mid-1980s, first-generation (1G) mobile networks were designed to support the wireless transmission of voice messages based on analog transmission with peak data rate up to 2.4 kbps. However, due to the absence of any universal standards, many drawbacks were emergent in the system, including poor security, lower transmission efficiency, and a problematical hand-off \cite{ars16}. 
Particularly, digital modulation technologies, including the Time Division Multiple Access (TDMA) and Code Division Multiple Access (CDMA) were employed to develop the 2G systems with better voice, short message services (SMS) and data rate of about 64kbps where the Global System for Mobile Communication (GMS) was used as the dominant mobile communication standard \cite{lxz18}. The high-speed data transmission 3G network was proposed during the year 2020 that provides quick access to the internet and a data transfer rate of about 2 Mbps thus, offering advanced services (i.e., web browsing, TV streaming, navigational maps, video services etc.) in comparison to the 1G and 2G networks \cite{sai}.
The IP based 4G network was introduced in the early 2000s to 1) improve the overall spectral efficiency, 2) decrease the latency, 3) provide high-speed downlink data rates of 1Gbits/s and uplink data rates of about 500Mbits/s, 4) accommodate Digital Video Broadcasting (DVB), High Definition TV content and video chat.  The 4G network offers terminal mobility (anywhere, anytime) through the automatic roaming across geographic boundaries of the wireless networks. The standards considered for the 4G networks include the Long Term Evolution-Advanced (LTE-A) and Wireless Interoperability for Microwave Access (WiMAX) \cite{pb11}.  LTE provides an integration of the existing and novel technologies, including coordinated multiple transmission/reception (CoMP), multiple-input multiple-output (MIMO), and orthogonal frequency division multiplexing (OFDM). 

After 4G, the commercial usability of the 5G would soon be enabled as the initial basic tests, the construction of hardware facilities, and procedures for standardization are almost complete. The 5G networks are basically aimed at the improvement in the data rates, reliability of networks, latency, efficiency in terms of energy, and enormous connectivity \cite{sms}. The 5G network offers higher data rates up to 10 Gbps due to the usage of a new spectrum of the microwave band (3.3-4.2 GHz) and involves advanced technologies such Beam Division Multiple Access (BDMA) and Filter Bank multi-carrier (FBMC). To improve the overall performance of the 5G network, numerous developing technologies including Massive MIMO (for capacity increase), Software Defined Networks (SDN) (for network flexibility), device-to-device (D2D) (for spectral efficiency), and Information-Centric Networking (ICN) is integrated into the network to allow rapid deployment of different services \cite{wdo}). However, for the 5G network, three scenarios of usage are proposed by the IMT 2020, including 1) Enhanced mobile broadband (eMBB), 2) Ultra-reliable and low latency communications (URLLC), and 3) massive machine-type communications (mMTC).
       
\subsection{Vision of 6G Networks}
\label{subsec_6G}
Various global research institutions have focused attention on the 6G networks as the 5G networks have entered the commercial deployment phase. The 6G networks are aimed at the enhancement of performance by the provision of peak data rates of about 1 Tbps and ultra-low latency (microseconds). Moreover, in comparison to the 5G networks, the 6G network is intended to improve the capacity by 1000 times through the usage of terahertz frequency and spatial multiplexing. The 6G networks will also provide global coverage through the effective integration of satellite and underwater communication networks \cite{ykk18}. Additionally, there are three novel classes for 6G networks, including the ubiquitous mobile ultra-broadband (uMUB), ultrahigh-speed-with-low-latency communications (uHSLLC), and ultrahigh data density (uHDD) \cite{zfw}.

\section{Marketing, Research Activities, and Trends Towards 6G Networks}       \label{sec_3}
\begin{figure}
\center
\includegraphics[width=3in]{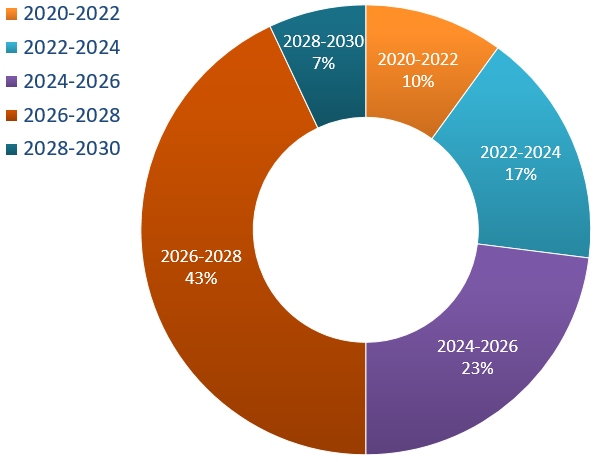} 
\vspace{2.5 mm}
\caption{Prediction of worldwide internet usage from 2020-2030 for consecutive 2 years (GB/month) }%\cite{r20}}
\label{Figure 3}
\vspace{-4 mm}
\end{figure}

\begin{figure}
\center
\includegraphics[width=3in]{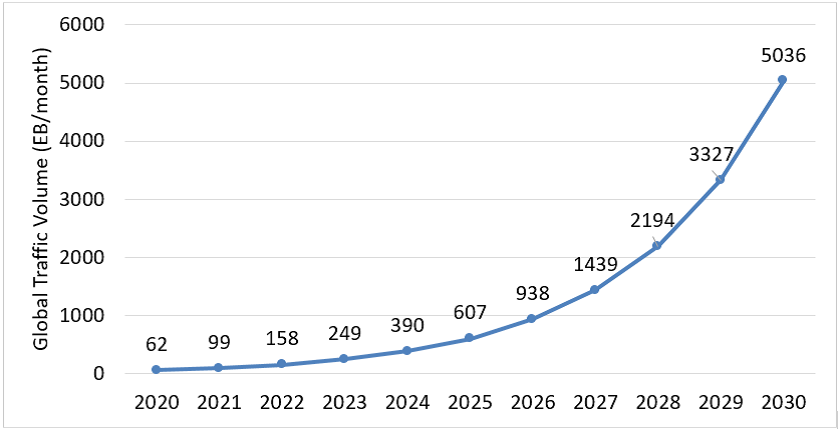} 
\vspace{2.5 mm}
\caption{The growth depicts worldwide connectivity during the years 2020-2030, in terms of the total global traffic volume}% \cite{jit, yxx, r20}.}
\label{Figure 4}
\vspace{-4 mm}
\end{figure}

\begin{figure}
\center
\includegraphics[width=3in]{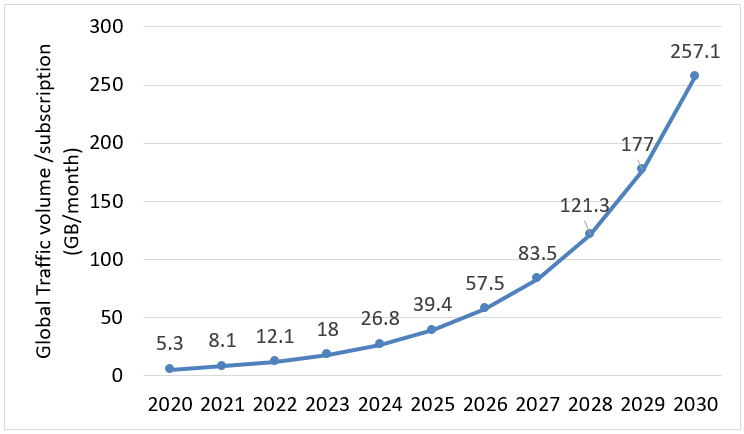} 
\vspace{2.5 mm}
\caption{The growth depicts worldwide connectivity during the years 2020-2030, in terms of traffic volume per subscription}% \cite{jit, yxx, r20}.}
\label{Figure 5}
\vspace{-4 mm}
\end{figure}

As far as the communication systems are concerned, a new generation is introduced every ten years since the first analog communication systems were introduced in 1980. Figure \ref{Figure 3} provides a representation of the worldwide internet usage (GB) that has increased considerably from 7\% (in the year 2020) to 43\% (in the year 2030) as a consequence of increased population from the year 2020 to 2030 \cite{r20}. The up-gradation from one generation to another brings along various improvements in the form of new services and new features where the goals of the 5G and 6G networks are to improve the overall capabilities of the networks through a factor of 10–100 in comparison to the previous mobile communication generations. However, during the last ten years, a phenomenal increase in mobile data traffic has been observed mainly due to the development and availability of smart devices and machine-to-machine (M2M) communications. The tremendous growth in the utility of mobile communications is reflected very well in Figure \ref{Figure 4} which depicts that in comparison to 2020, the expected worldwide mobile traffic volume will increase 700 times in the year 2030 \cite{jit}. Moreover, it is predicted by the International Telecommunication Union (ITU) that the overall mobile data traffic will prominently exceed 5 ZB per month and the number of mobile subscriptions will reach 17.1 billion by the end of the year 2030 as shown in Figure \ref{Figure 5} \cite{r20}.

It is anticipated that the annual growth rate of approximately 70\% will be evident for the 6G network from the years 2015 to 2030 subsequently, reaching a value of 4.1 billion US dollars by the year 2030 \cite{kyi}. Since the 6G networks have various advanced communication infrastructures, including edge computing, cloud computing, and AI, they will ultimately offer greater market shares, i.e., up to 1 billion US dollars \cite{chs}. AI-based chipsets are another major component of the 6G networks that will rise above 240 million units by the year 2028. Different worldwide organizations have started extensive research projects on the 6G mobile communication networks \cite{kyt}. One of the most important research programs is the 6G Flagship research program that was supported by various working bodies, including the Academy of Finland VTT Technical Research Center, Oulu University of Applied Sciences, Nokia, Business Oulu,  Aalto University, InterDigital, and Keysight Technologies \cite{krz}.

%Table 3
% \usepackage{booktabs}
% \usepackage{graphicx}
\begin{table*}[]
\centering
\caption{A detailed comparison of 6G with the previous mobile communication technologies }%\cite{chs}}
\label{Table 3}
\resizebox{\textwidth}{!}{%
\begin{tabular}{@{}lllllll@{}}
\toprule
Specifications             & 1G          & 2G         & 3G     & 4G            & 5G           & 6G                  \\ \midrule
Data rate                  & 2.4 kbps    & 64 kbps    & 2 Mbps & 100-1000 Mbps & $\approx$ 20 GBPS    & $\approx$ 1 TBPS            \\
End-to-end latency         & 20- 200 sec & 10-100 sec & 1 Sec  & 100 ms        & 10 ms        & 1 ms                \\
Highest spectral efficiency & 1 bps/ HZ      & 0.5 bps/ HZ     & 2.5 bps/Hz       & 15 bps/Hz        & 30 bps/Hz        & 100 bps/Hz        \\
Network mobility support    & Up to 15m/hour & Up to 50km/hour & Up to 150km/hour & Up to 350km/hour & Up to 500km/hour & Up to 1000km/hour \\
fmax                       & -           & -          & -      & 5 GHz         & 90 GHz       & 10 THz              \\
XR                         & -           & -          & -      & NO            & Partial      & Full                \\
THZ communication          & -           & -          & -      & NO            & Very Limited & Wide                \\
Services                   & -           & -          & -      & Video         & VR, AR       & Tactile             \\
System Architecture        & -           & -          & -      & MIMO          & Massive MIMO & Intelligent surface \\
AI                         & NO          & NO         & NO     & NO            & Partial      & Full                \\
Autonomous vehicle         & NO          & NO         & NO     & NO            & Partial      & Full                \\
ER (Extreme Reality)       & NO          & NO         & NO     & NO            & Partial      & Full                \\
Haptic Communication       & NO          & NO         & NO     & NO            & Partial      & Full                \\
SI (Satellite integration) & NO          & NO         & NO     & NO            & NO           & Full                \\ \bottomrule
\end{tabular}%
}
\end{table*}

The Flagship research program for 6G was initially carried out to co-create an ecosystem for innovation in 6G and adopt 5G networks. The basic aim behind the 6G Flagship research program is to develop a society that is driven through unlimited and high-speed wireless connectivity. Additionally, to streamline the development of the 6G technology South Korean government signed an agreement with the University of Oulu, Finland \cite{lz20}. To carry out the 6G network-based research, LG has also established a research laboratory at the Korea Advanced Institute of Science and Technology \cite{ln20}.   SK Telecom, with other partners including Samsung, Nokia, and Ericsson, also initiated a joint research project on 6G-based technologies  \cite{kyt}.
Moreover, 6G-based research activities have also been initiated in China, and Huawei has already begun research on the 6G networks at its Ottawa-based research center in Canada \cite{zxm, hl20}. Most prominently at the NYU WIRELESS research center, several faculty members are actively involved in research on various core components of the 6G networks, including machine learning, quantum nano-devices, communication foundations,  and 6G testbeds \cite{akn20}. Last but not least, the US has also announced an active investigation of the 6G networks by initiating numerous 6G-based research programs \cite{h20}.

 \section{6G Networks Requirements}
In recent years, many studies focused on the 6G applications, facilitating technologies, architectures, and open research challenges have been reported extensively in the literature. Towards this end, \cite{sbc19} introduced applications, facilitating technologies, and some open research challenges for the 6G technology. Moreover, they also addressed applications, performance metrics, 6G driving trends, as well as new customer services for 6G networks \cite{sbc19}. The concept of AI endowed 6G wireless networks was introduced in \cite{lcs}. The 6G network design and the applications of 6G for different AI-empowered smart services were also elaborated \cite{lcs}. Tariq et al. concentrated on the use of 6G, its enabling technologies, as well as research challenges \cite{tkw}. Moreover, Giordani et al. highlighted the development of wireless communication systems on the way to 6G networks and also some of its use cases \cite{gpm}. They discussed mainly the key enabling technologies of 6G, their related challenges, and their applications. Reference \cite{gpm} also presented the concept of intelligence integration in 6G systems.

Furthermore, the key drivers, requirements, design, and enabling technologies of 6G are discussed in \cite{zfw}.  The potential technologies for 6G wireless networks have also been elucidated in \cite{yxx}. They presented a summary of time, frequency, space, and resource usage relevant to the 6G networks.  Moreover, important techniques involved in the evolution of 6G wireless networks and the upcoming problems concerning the implementation of 6G  were also elaborated in \cite{yxx}.  Additionally, the peak data rate, energy efficiency, connectivity density, user experienced data rates, as well as latency of 6G is discussed in \cite{zln19}. Moreover, the challenges of 6G wireless system concerning its intelligence and various machine learning schemes are presented in \cite{kmt}. Akyildiz et al. provided a comprehensive discussion on key enabling technologies of 6G \cite{ajh14}. Moreover, intelligent communication conditions with their layered structural design and open research challenges are discussed in \cite{akn20}. The technology trends in 6G, its applications, the requirements, and the concept of 6G are discussed in \cite{cls}. A detailed analysis of the existing 6G-based research studies is provided in Table \ref{Table 4}. In the following, we discuss some of the key requirements of 6G networks.

\begin{table*} [!]
\renewcommand{\arraystretch}{}
\noindent\begin{minipage}{\linewidth}
\centering
\captionof{table}{A Comprehensive survey on 6G communication networks}\begin{tabular}{ p{5cm} p{12cm}  }
\label{Table 4}
\\\hline
Author  Year                    & Main Objective Of Research                                                                                                                                                                                                                                                            % & Area To Be Focused                            & Limitation                                                  
\\ \hline
Nawaz et al., 2019 \cite{nsm}            & Presented a comprehensive review on B5G applications, issues, use cases and potential benefits of quantum computing and machine learning.                                                                                                                                              %& Quantum computing and machine learning        & Practical implementation                                    
\\
                         %       &                                                                                                                                                                                                                                                                                        &                                               &                                                             \\
Saad et al., 2019 \cite{sbc19}             & A comprehensive performance requirement of 6G technology and its proposed application trends are elaborated.                                                                                                                                                                         %  & 6G performance                                & Spatial spectral efficiency                                 
\\
                        %        &                                                                                                                                                                                                                                                                                        & requirement                                   &                                                             \\
Strinati et al., 2019\cite{sbg}        & Analysis the research gap of previously used technology and predicting the 6G roadmap for future communication.                                                                                                                                                                       % & Predicting 6G roadmap                         & Resource limitation of edge devices                         
\\
Salehi and Hossain 2019\cite{sh19}       & The challenges associated to UAV network related to temporal correlation for distribution and success probability is presented.                                                                                                                                                       % & UAV network                                   & Proper communication range                                  
\\
Huang et al., 2019\cite{hyw}           & Wireless communication technology that provides solutions for the bottlenecks that limit the capability of the integrated space and terrestrial network (ISTN) are proposed.                                                                                                          % & ISTN                                          & Power limitation because of high power terahertz radiation  
\\
Elliott et al., 2019\cite{ekm19}          & Future cellular networks and short range communication is discussed.                                                                                                                                                                                                                  % & Automated vehicles                            & Long range or aerial communication                          
\\
Ji et al., 2019\cite{jlz}               & A detailed survey on 5G/B5G wireless communication for UAVs.                                                                                                                                                                                                                           %& UAV wireless communication                    & Energy requirement                                          
\\
Letaief et al., 2019\cite{lcs}          & A comprehensive discussion on AI enabled 6G applications and optimized network architecture using state-of-the-art technologies.                                                                                                                                                    %   & 6G for AI                                     & Computation time and accuracy                               
\\
Yang et al., 2019\cite{yxx}             & 6G techniques and future research trends to improve it are analyzed.                                                                                                                                                                                                                %   & Improving 6G technology                       & Open research problems                                      
\\
                         %       &                                                                                                                                                                                                                                  %                                                      &                                               &                                                             \\
Chowdhury et al., 2019\cite{chh}        & A detailed overview on 6G enabled AI wireless communication                                                                                                                                                                                                                            %& 6G for AI                                     & Computation time and accuracy                               
\\
Zhang et al., 2019\cite{zln19}           & Incorporating the three main aspects, AI, IoT and mobile ultra-broadband for evolving 6G technology.                                                                                                                                                                           %        & AI, IoT and mobile ultra-broadband            & Mobile data coverage                                        
\\
Lovén et al., 2019\cite{llp}            & 6G wireless communication and role of Edge AI is elaborated for future.                                                                                                                                                                                                               % & 6G for AI                                     & Computation time and accuracy                               
\\
Clazzer et al., 2019\cite{cml}          & Recent advances in modem random access and uncoordinated medium access for different IoT applications in 6G paradigm is discussed.                                                                                                                                                    % & 6G enabled IoT applications                   & Network design and optimization                             
\\
Giordani et al., 2020\cite{gpm}         & 6G use cases and their requirements are presented.                                                                                                                                                                                                                                     %& Use cases for 6G                              & Designing of a energy efficient system                      \\
                               % &                                                                                                                                                                                                                                %                                                        &                                               &                                                            
\\
Viswanathan and Mogensen 2020\cite{vm20} & A detailed overview and performance requirement for 6G technology transformation is discussed. Highlighting the privacy issues, latency, reliability, sensing capabilities, spectrum bands, network architecture and spectrum methods.                                     %            & 6G technology                                 & Anonymizing information                                     \\
                            %    &                                                                                                                                                                                                                                                                                        & transformation                                &                                                             
\\
Tariq et al., 2020\cite{tkw}            & Extending the vision of 5G to provide step changes for enabling 6G.                                                                                                                                                                                                                   % & Use cases for 6G                              & Proper infrastructure                                       
\\
Mahmood et al., 2020\cite{mal20}          & Different machine type communications, trending technologies and performance indicators for 6G are discussed.                                                                                                                                                                      %    & 6G machine type                               & Low power wide area                                         \\
%.                               &                                                                                                                                                                                                                                                                                        & communication                                 &                                                             
\\
Dang et al., 2020\cite{das}             & Explores the Challenges associated with 6G deployment and a future vision is incorporated.                                                                                                                                                                                         %    & Privacy, security and applications            & Energy efficient designing                                  
\\
Zhang et al., 2020\cite{zll}            & Wireless evolution towards 6G communication is surveyed. Enhanced network architecture, ubiquitous 3D coverage, protocol and persuasive AI is highlighted.                                                                                                                          %   & Ubiquitous wireless communication             & Network architecture                                        
\\
Zhang et al., 2020\cite{zdw}            & Categorizing the current technologies and extending the drive force by AI-enabled intelligent communication.                                                                                                                                                                        %   & AI-enabled intelligent communication.         & 6G integrated with AI technology                            
\\
Gui et al., 2020\cite{glt}              & 6G requirements are achieved using the five 6G core components. Additionally, how to enable KPIs and centricities are discussed in detail to address these components.                                                                                                                % & Use cases for 6G                              & Designing of a energy efficient system                      
\\
Zhang et al., 2019\cite{zxm}           & Incorporating the three main aspects, AI, IoT and mobile ultra-broadband for evolving 6G technology.                                                                                                                                                                            %       & AI, IoT and mobile ultra-broadband            & Mobile data coverage                                        
\\
Tomkos et al., 2020\cite{tkp}           & A comprehensive overview of the transformation of IoT technologies towards 6G networks is presented.                                                                                                                %              & Edge computing                          %      & Security challenges and cost                                
\\
Yaacoub and Alouini 2020\cite{ya20}      & A survey on connectivity for rural areas is presented. Additionally, backhaul and front haul techniques are analyzed using cost efficiency and energy requirements.                                                                                                 %                   & backhaul and front haul techniques            & Power limitation on transmission                            
\\
Kato et al., 2020\cite{kmt}             & The IoT networks for 6G are discussed, where the IoT devices are connected using different frequency bands, such as mmWave and THz                                                                                                                                                     %& 6G IOT networks                               & Energy efficient                                            
\\
Shafin et al., 2020\cite{slc}           & A comprehensive overview on the applications, challenges and future research direction for B5G and 6G networks are presented.                                                                                                                                                       %   & B5G and 6G networks                           & Privacy issues                                              
\\
Gui et al., 2020\cite{glt}              & A survey on the machine learning techniques for network, security and communication of 6G vehicular technology is presented.                                                                                                                                                    %       & 6G vehicular technology                       & Long range communication                                    
\\
Zhang et al., 2020\cite{zdw}            & Low latency networks are supported using reinforcement learning framework and a heterogeneous multi-layer edge computing is presented.                                                                                                                                                % & Heterogeneous multi-layer edge computing      & Efficient resource utilization                              
\\
Chowdhury et al., 2020\cite{csa}        & A comprehensive literature review on the probable 6G technologies, the requirements, applications and technologies that are expected to evolve in the near future for the 6G networks. Moreover, the associated challenges with these emerging technologies have also been elaborated.% & Emerging 6G technologies requirements, use cases, network architecture, future challenges, and future directions for research            &                                                             
\\
Kim 2021\cite{k21}                      & Provides an elaboration of the main components, description of enabling technologies, current  research and possible applications of the 6G wireless communication systems to IoT based services/technologies                                                                         % & 6G and IoT                                    &                                                             
\\
Allam and Jones 2021\cite{aj21}          & Presents the scope and emerging directions for the 6G applicability to the Digital Twins and Immersive Realities. Provides an extensive overview of 6G, associated concepts, and relations in context of the future Smart, Digital and Sustainable Cities                  %            & 6G role in smart city                         &                                                            
\\
Padhi and Charrua-Santos 2021\cite{pc21} & The synthesis of 6G, IoT, IoE, industrial Internet of Everything (IIoE) is presented here. This study also reports a novel theoretical framework for 6G-enabled IIoE (6GIIoE) system.                                                                                             %     & theoretical framework for the 6G-enabled IIoE &                                                             
\\
Yang et al., 2021\cite{ycw}             & It embodies the key requirements for the application of federated learning (FL) to the wireless communication systems.                                                                                                                                     %                            & Federated Learning for 6G                     &                                                             \\
                       %         &                                                                            &                                               &                                                             
\\
Wang et al., 2021\cite{wth}             & A security scheme based on the Internet-of-Vehicles (IoV) devices that request services from the edge nodes anonymously is presented here.                                           %                                                                                                  & ~Block-Streaming Service Awareness            &                                                             
\\
Shahraki et al., 2021\cite{sap}         & ~The article highlight the importance of 6G networks, its requirements, major trends, latest research, performance indicators, and applications relevant to 6G networks. Moreover, the study provides the depiction of various unresolved challenges for the future utility of the 6G.% & Applications of 6G                            &                                                             
\\
Imoize et al., 2021\cite{iat}           & The enabling technologies, emerging 6G applications, technology mediated challenges, possible solutions and other issues (social, psychological, commercialization) relevant to the vision of 6G are elaborated in detail.                                                         %    & Smart infrastructure of 6G                    &                                                             
\\
Wang 2021\cite{w21}                     & The application scenarios of data mining (in subjects/contents) for online teaching (quality control) based on the 6G networks are described here.                                                                                                                                    % & Online teaching based on 6G                   &                                                        
\\
\\ \hline
\end{tabular}
\end{minipage}
\end{table*}

%\subsection{Rationale for 6G Networks for Disaster Management}
\subsection{Connectivity}
In the near future, societies will become potentially data-driven through the utilization of prompt and unlimited wireless connections \cite{db18}. Generally, to allow the 5G network utility for various smart applications different approaches including, new 5G radio, the simultaneous usage of unlicensed and licensed bands are brought into consideration \cite{ars16, hcz18, lfz18}.  The anticipated benefits of the 5G networks in the form of basic smart IoE-based services and short packets for URLLC display inherent liabilities and complexities to completely fulfill the requirements of the future smart city IoE applications \cite{sbc19, ywh}. Thus, it is evident that the capabilities and important performance indicators of the 5G network are not adequate to meet the increased requirements arising from the development of different data-centric and automated processes \cite{gpm}. The applications of telemedicine, haptics, and connected autonomous vehicles, are envisioned to utilize long packets with ultra-high reliability and high data rates, thus violating the general usability of short packets for URLLC that are implemented by the 5G networks \cite{sbc19}. Another limitation of the 5G network in terms of the exceeding demands of next-generation smart industries is the unsuitable connectivity density of 106/$\text{km}^2$ \cite{y15}. Some of the major shortcomings of the 5G networks are the short mmWave connectivity range, Gbps level transmission data rate, the interruptions in the signals, and no/limited coverage for the rural/remote areas \cite{ywh}.

\subsection{Latency}
Low latency, or the deterministic latency that requires the use of deterministic networking (DetNet), is one of the distinguishing features of 5G networks that are used to assure timely and accurate end-to-end latency. By far, 6G mobile networks will offer additional facilities such as high time and phase synchronization accuracy better than that offered by the 5G networks \cite{ywh}. Therefore, it is well established that 6G will emerge as a promising technology that will meet the requirements of various diverse sectors to improve the quality and perception of life in the near future \cite{np21}. Not limited to this, the 6G networks will effectively overcome the limitations of 5G networks, also catering to the exceeding requirements of the next-generation smart systems. In comparison to the 5G, the 6G networks are intended to provide promising features such as much higher spectral/energy/cost efficiency, nearly 100\% geographical coverage, 10 times lower latency, sub-centimeter geo-location accuracy, millisecond geo-location update rate, high-level intelligence for full automation, sub-millisecond time synchronization, a higher transmission data rate (Tbps), and a connection density that is 100 times higher \cite{y15, ajh14}.

%\subsection{Rationale for 6G Networks in Smart Cities}
\subsection{Reliability}
The 6G networks are expected to provide 99.9\% reliability \cite{kgt21}. Moreover, 6G will use artificial intelligence (AI) as an integral part which will prove beneficial for the optimization of a wide array of wireless network problems \cite{asr}. The deployment of 5G networks has provided a realization of the fact that softwarization pays a cost as the usage of the commercial off-the-shelf (COTS) servers instead of the domain-specific chips in a virtualized radio access network (RAN) implicates a large increase in energy consumption thus, requiring measures for improving the energy efficiency. This can be explained by the fact that in comparison to the 4G networks, the 5G networks deliver a higher bandwidth at the cost of higher power consumption. Therefore, it is of extreme importance for the 6G networks to require a relatively new computing paradigm that should leverage the benefits of softwarization without paying the costs in terms of energy consumption \cite{y15}. Moreover, it is well established that most of the 6G use cases will eventually evolve from the emerging functionalities and quality of experiences of the 5G system-based applications. The applications of the 6G networks will proceed further by the performance enhancement measures, and the addition of new use cases \cite{tkw}. The details on the use cases for the 5G and 6G networks are provided for comparison in Table \ref{Table 5}.

\subsection{Computing Techniques}
Important computing technologies, including cloud computing, fog computing, and edge computing, form an integral part of distributed computing, processing, lower latency, synchronization time, and overall network resilience. In addition to short-packet drawback, it is highly anticipated to overcome other limitations of the 5G networks through the provision of higher reliability, lower latency, better system coverage, and higher data rates \cite{sbc19}. Moreover, the 6G should be based on a human-centric approach rather than the machine-,  application- or data-centric approaches to meet the mobile communication demands of the coming years  \cite{das, zfw}.    

\subsection{Coverage}
\begin{figure*}[t]
\center
\includegraphics[width=\linewidth]{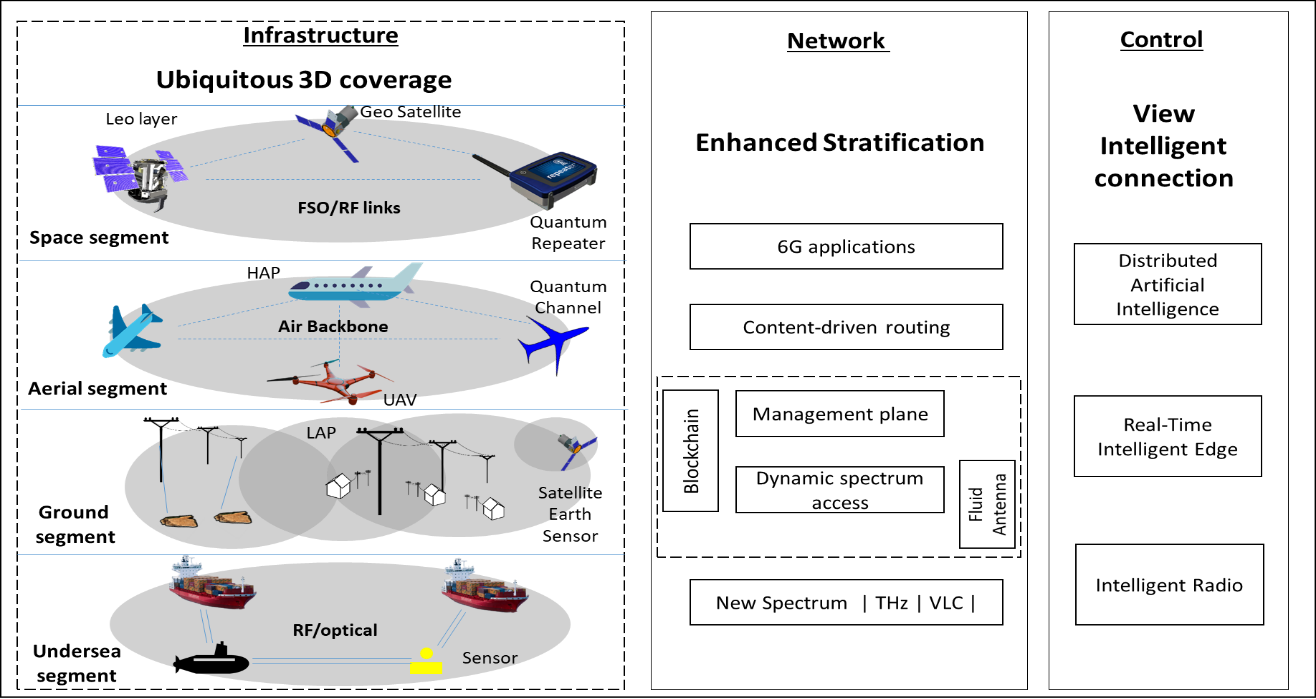} 
% \vspace{2.5 mm}
\caption{Overview of the 6G architecture.}% (Huang et al., 2019).}
\label{Figure 6}
% \vspace{-4 mm}
\end{figure*}

The elucidation of new paradigm shifts will provide the essence of the 6G wireless networks. The 6G networks will provide global coverage of the integrated networks of the space, ground, air, and sea. The overview of the 6G architecture is presented in Figure \ref{Figure 6} \cite{hyw}. The coverage and range of the wireless communication networks can be extended extensively through the usage of satellite communication, UAVs, and maritime communication \cite{saeed2020cubesat}. 

\subsection{Data Rate}
The overall improvement in the data rate can be enabled by exploring all spectra, i.e., optical frequency bands, sub-6 GHz, mmWave, and THz. Additionally, the utility of Artificial Intelligence and Machine learning techniques in combination with the 6G networks would ultimately allow the full applicability, automation, and network management of the 6G. AI-based approaches can significantly improve the next-generation network performance by providing dynamic instrumentation of the networking, caching, and computing resources.

\subsection{Security}
A stronger network security needs to be implemented during the development procedure for both physical and network layers in 6G. Last but not least, the development of the 6G networks will be boosted considerably through the utilization of industry verticals, including cloud VR, IoT, industry automation, cellular vehicle to everything (C-V2X), area network for the digital twin body, and energy-efficient wireless network control and federated systems of learning \cite{y15}. Therefore, security would be of paramount importance in 6G systems.

\begin{table*}[]
\centering
\caption{Details on the use cases for the comparative analysis of the 5G and 6G networks\cite{tkw, vm20}.}
\label{Table 5}
\renewcommand{\arraystretch}{1.3}
\resizebox{\textwidth}{!}{%
\begin{tabular}{ p{6cm} p{6cm} p{6cm} }
\toprule
Use Case                                                & 5G                               & 6G                \\ \midrule
Centre of gravity                                       & User-centric                     & Service-centric   \\
Augmented reality for industry in terms of Peak rate and capacity &
  Low resolution and high level tasks &
  High resolution with multi sensing and comprehensive level tasks \\
Tele-presence in terms of capacity &
  Limited scale and a high video quality &
  Mixed reality \\
Security surveillance, detection of defects in terms of positioning and sensing &
  External sensing with limited automation &
  Fully automated through the integrated radio sensing \\
Dynamic digital twins and virtual worlds                & No                               & Yes               \\
Data center wireless in terms of capacity and peak rate & No                               & Yes               \\
Automation, distributed computing in terms of time synchronization &
  Micro second level tasks &
  High precision tasks at nano second level \\
Ultra-sensitive applications                            & Not feasible                     & Feasible          \\
Zero energy devices                                     & No                               & Yes               \\
Groups of robots or drones in terms of  low latency     & Might be                         & `Yes              \\
Bio-sensors and AI                                      & Limited                          & Yes               \\
True AI                                                 & Absent                           & Present           \\
Reliability                                             & Not extreme                      & Extreme           \\
VAR                                                     & Partial                          & Massive scale     \\
Time buffer                                             & Not real-time                    & Real-time         \\
Capacity                                                & 1-D (bps/Hz) or 2-D (bps/Hz/m2 ) & 3-D (bps/Hz/m3  ) \\
VLC                                                     & No                               & Yes               \\
Satellite integration                                   & No                               & Yes               \\
WPT                                                     & No                               & Yes               \\
Smart city components                                   & Separate                         & Integrated        \\
Autonomous V2X                                          & Partially                        & Fully             \\ \bottomrule
\end{tabular}%
}
\end{table*}

\section{Essential Enabling Technologies for 6G Networks}
The evolution of the mobile networks is based on inheriting the advantages of the previous network architectures and adding extra benefits that effectively meet the requirements of the latest era\cite{yxx}. Similarly, the 6G network will adopt the benefits of the 5G architectures also concurrently new technologies to overcome the future demands, thus, it is indicated that the 6G communication systems will be mediated by various technologies some of which are discussed as follows:
\subsection{Internet of Things (IoT)}
IoT seeks to connect everything to the Internet, establishing a connected environment where data sensing, processing, and communications are conducted automatically without human participation, as a major technology in integrating heterogeneous electronic devices with wireless networks. End users can benefit from IoT data acquired from ubiquitous mobile devices including sensors, actuators, smart phones, computers, and radio frequency identifications (RFIDs) \cite{8907365}. According to Cisco \cite{9512099}, by 2030, up to 500 billion IoT devices will be connected to the Internet. In addition, according to a new study by IHS Markit \cite{9512098}, a worldwide leader in critical information, analytics, and solutions, the number of linked global IoT devices will grow at a staggering 12 \% per year, from roughly 27 billion in 2017 to 125 billion in 2030. 

In this perspective, 6G will be a significant enabler for future IoT networks and applications, as it will provide full-dimensional wireless coverage and integrate all functionality, including sensing, transmission, computation, cognition, and fully automated control. In fact, compared to the 5G mobile network, the next generation 6G mobile network is expected to give massive coverage and enhanced adaptability to support IoT connectivity and service delivery \cite{7397856}.

\subsection{Artificial Intelligence (AI)}
One most crucial component of the self-sufficient 6G networks is intelligence, which is a relatively new technology being integrated in the 6G networks through the utility of AI \cite{9524496,sa19, z19}. It is evident that AI could not be applied to the previous versions including 4G and lower generation. However, in the 5G networks a partial or limited applicability of AI will be observed. Most prominently the 6G networks would provide the full automation through AI which will offer full potential of the radio signals, also allowing the cognitive radio to intelligent radio based transformations \cite{lcs}. It is notable that for 6G real time communications the advancements in the machine learning/AI procedures leads to the development of highly intelligent networks that will ultimately improve and simplify the real-time data transmission. AI techniques display numerous benefits such as increasing efficiency, reducing the processing delays within the communication steps, solving complex problem efficiently, prompting communications within the BCI, performing network selection and handover. However, instances such as meta-materials, intelligent structures, intelligent networks, intelligent devices, intelligent cognitive radio, self-sustaining wireless networks, and machine learning would provide the support for the communication systems based on AI \cite{mal, z19, chs}. 

Therefore, the application of AI based technology will assist in meeting the goals of several 6G services including uMUB, uHSLLC, mMTC, and uHDD. The recent advancements in the machine learning approaches allow its application to RF signal processing, spectrum mining, and spectrum mapping. Whereby the combination of machine learning approaches with photonic technologies will also uplift the AI evolution in 6G networks to shape a cognitive radio system that is based on photonics. For the channel state estimation, and automatic modulation classification, the physical layer implements the AI based deep learning encoder-decoder based setup, whereas, deep learning-based resource allocation, intelligent traffic prediction and control have been extensively investigated for the data link layer and transport layer respectively. An additional advantage associated to the application of machine learning and big data is the determination of best possible approaches for data transmission between the end users through the provision of the predictive analysis \cite{mal, z19, chs}. 
\subsection{Integration of Wireless Information and Energy Transfer}
One of the most ground-breaking technologies within the 6G network is the integration of the Wireless Information and Energy Transfer (WIET) which takes in to account a set of fields and waves that are similar to those used in wireless communications. Since, WIET shows a greater potential for lengthening the battery charging lifetime of the wireless systems, thus, providing support to the devices without batteries in the 6G networks \cite{whg}. WIET is particularly envisioned to allow the progression of battery-less smart devices, charging and saving the battery life-time of the wireless networks and other devices respectively \cite{jkc15, es20}.

\subsection{Mobile Edge Computing}
The launching of content delivery networks (CDNs) in 1990s by Akamai is the first step towards edge computing for performance and speed improvement. Moreover, edge computing takes a broad view of CDN concept by utilizing cloud computing platform. Brain and his co-workers in 1997 introduced the importance of edge computing to mobile networks \cite{mba20}. However, the cloud computing began to rise in mid 2000s and became the most usable infrastructure for mobile devices which is used today by Apple and google devices. Paramir Bahl and his colleagues were first to demonstrate the conceptual groundwork of edge computing in 2009 \cite{gei}. Edge computing is of great importance by creating new onsets in computing environment. It allows the services of cloud computing to come closer to end user as it is the modified version of cloud computing and lessens the delay time of bringing services to the end user. It is a fast-processing system that has very quick response time \cite{krz}. The upcoming 6G networks will integrate the current 5G and IoT infrastructures, with the help of the edge computing hardware, thus, supporting the heavy execution of AI algorithms \cite{tkp}. 

Therefore, the mobility enhanced edge computing (MEEC) will become an integral part of the future 6G machinery due to immense applications of the distributed large scale clouds. Moreover, the amalgamation of MEC infrastructures with AI methods will allow effective computation not only on the big data analytic but also on the system controls to the edge. Edge based intelligent computing has emerged to leverage maximum benefits in fulfilling the challenging needs of the impending heterogeneous computation, communication, and high-dimensional intelligent configurations based ubiquitous service scenarios\cite{cls}. Various applications of the edge computing are illustrated in Figure \ref{Figure 7}. Herein, various applications have been considered to highlight the importance of edge computing. The applications are elaborated as follows:

\begin{figure}
\center
\includegraphics[width=2.9in, height= 6cm]{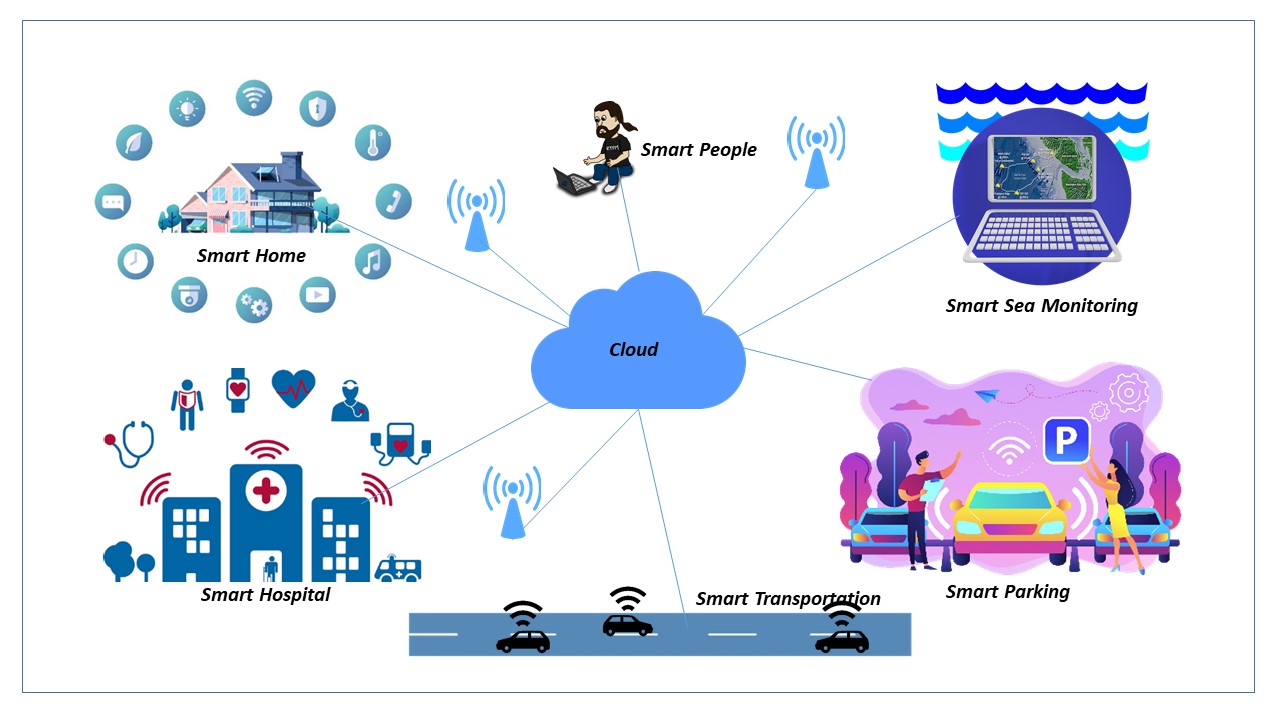} 
\vspace{2.5 mm}
\caption{Applications of edge computing.}
\label{Figure 7}
\vspace{-4 mm}
\end{figure}

\subsubsection{Real-Time Reporting of Autonomous Cars Accidents}
The future of cars will be holding a huge number of autonomous cars. There are six levels of autonomous cars including level 0,1,2,3,4,5. The difference in levels of car is their varying automation levels i.e., level zero has no automation whereas level 5 represents full level of automation in cars. Moreover, these cars have the ability to handle lane changing and also to tackle the collision \cite{llt}.  The roadside units having edge computing facility can be fruitful for handling real time data of such cars. Considering the example of road accident where the timely reporting of accident is linked with providing the first aid to the patients. In that case the key factors are the medical team along with the road administrators, the smart police and public safety points having edge computing facility for effective communication.  During the emergency situation the damage can be minimized by providing the timely first aid and this can be achieved by reporting the incident at an appropriate time.  Hence, these incidents are either reported by the person himself or with the help of advanced edge computing enabled safety points. It is understood that the person might have severe injuries and be unable to report about accident which can lead to an adverse situation.  Therefore, the safety points having edge computing facility will automatically detect the incident by applying algorithms and result in timely detection of the incidents.  Thus, edge computing is necessary for detection of such delicate tasks \cite{kyi}.
\subsubsection{Smart Forest Fire Detection}
The concept of smart forest originated from the IoT. In this concept the data about environmental conditions is collected via remote sensing.  The foremost objective of this detection is to control wild fire in the forest at the near beginning phase. Smart forest fire detection system might decrease the harm caused by the fire in forests. The smart forest fire detection based on edge computing facility has assisted in timely reporting the forest fires, hence can be effectively used for monitoring of fire \cite{kad18}. Here, cameras are fitted in the cars that are taking pictures constantly. Hence, fires are reported by processing the pictures to the server. The delay in processing of picture may result in loss of communication leading towards late response against fire. Thus, edge computing based on image processing will help reduce this delay and as a result quick decisions could be made against the fires due to timely reporting \cite{nae18}. Moreover, it also shows applications in several rescue activities via the telecommunication process.
\subsubsection{Smart Parking System }
All activities in the day-to-day life of urban cities like working, shopping etc demands the parking at an inexpensive place. So, using the internet connections users can search for vacant parking slots. Conventional parking system frequently faces the challenge of finding the vacant parking space and also inefficiency in management of parking space. Hence, smart parking systems have been introduced to solve these problems. These systems use various machine learning algorithms for rapid computation of vacant parking space and also provides the systematic management of parking space \cite{blk}. The image recognition system enabled by edge computing helps in facilitating the smart parking systems. These image recognition systems find the vacant spaces for parking using various AI algorithms \cite{blk}. Therefore, to avoid these inconsistencies efficient and smart parking system must be designed for empty space detection in a very small time. Thus, for that purpose edge computing will be helpful in enabling smart parking systems \cite{kzp}. 

\subsubsection{Smart Home}
Currently, an enormous amount of services are installed on the edge of the network from the cloud due to the fact that processing data at the edge can decrease response time and lower throughput costs for applications such as systems in smart homes in order to help improve the living comfort of residents \cite{clx18}.  Various features are designed for smart homes such as improved surveillance, smart controls, and smart meters. However, for the implementation of such smart home ideas a multi-layer system is demanded that is able to make decisions about home automation \cite{cd17}.  Hence, different AI algorithms as well as IoT devices are brought in to account to for the multi-layer system to carry out its functions. The system uses real-time and historical information to make decisions. There are various designed devices that are already using such systems such as smart TV, refrigerators, air conditions, washing machines etc. Hence, edge computing is really helpful in implementation of such smart hope ideas using artificial intelligence algorithms \cite{scz}. Therefore, vision of 6G networks will provide a strong base for the implementation and execution of various smart services.

\subsection{Integration of Sensing and Communication}
One principle mediator of the self-directed wireless networks is the ability to detect the dynamically changing environmental states and allowing effective exchange of information among various nodes \cite{kck18}. The autonomous or the self-directed systems would be supported in the 6G networks through the effective integration of sensing and communication using a large number of sensing objects, complex resources for communication, computing resources (multi-level) and cache resources (multi-level) which is a challenging endeavor \cite{chs}. 

\subsection{Dynamic Network Slicing }
The dynamic network slicing is an important aspect that needs to be considered by the operators of a network to ensure dedicated execution of the virtual networks and to provide extensive support for the optimized delivery of services towards various users (i.e., vehicles, industries, machines). Thus, dynamic network slicing serves as a core element in the 5GB communication systems mainly for the management of multiple users that are connected to numerous heterogeneous networks. As far as the implementation of the dynamic network slicing is considered, it requires software defined networking and network function virtualization techniques. These techniques are influential for the cloud computing in the management of networks for the performance optimization through a centrally controlled dynamic steering, traffic flow management and organized network resource allocation \cite{chs, sgw}.
\subsection{Holographic Beamforming}
The beamforming procedure is based on the signal processing through which the radio signals can be transmitted in precise directions using an array of steered antennas through an emphasis on the minimized angular range \cite{es20}. Moreover, beamforming procedure displays a wider range of benefits in the form of higher network efficiency, better coverage and throughput, higher signal to interference noise ratio (SINR), user tracking, interference prevention and rejection \cite{chh, es20}. Holographic beamforming (HBF) is a new method that is based on the usage of a hologram to achieve steering of the beam through the antenna where, the RF signals emerging from the radio travel to the back of the antenna thus, scattering across its front and later adjusting according to the beam shape and direction \cite{es20}. This method is based on the utility of Software Defined Antennas (SDA) hence, it is quite different from MIMO systems because in comparison to the traditional arrays or MIMO systems the SDA are smaller, lighter, cheaper and low power consuming \cite{b17}.  Since Cost, Size, Weight and Power (C-SWaP) are the major challenges for the design of any communication system but the utilization of SDAs in the HBF procedure will lead to the highly flexible and efficient transmission and reception of signal in the 6G networks \cite{es20}. Particularly, for the 6G networks the use of HBF approach in multi-antenna communication devices is advantageous for the transmission and reception of signals in a highly efficient and flexible manner. Thus, indicating important roles of HBF positioning in the wireless power transfer, physical layer security and augmented network coverage related scenarios\cite{chh}. 
\subsection{Big Data Analytics}
The analytics of Big data is a highly intricate and complicated procedure, that is used for the analysis of a broader range of massive data sets through revelation of information associated to concealed pattern, unidentified correlations, and customer dispositions to guarantee data management in comprehensive manner. The collection of Big data occurs from a variety of sources (i.e., videos, social networks, images, sensors). Moreover, the Big Data Analytics is effectively deployed to handle and manage massive amounts of data within the 6G communication networks. The deployment of large amounts of data, deep learning protocols and big data analytics within the 6G networks are foreseen to lead to advancements in the 6G network mainly because of the automation and self-optimization properties. End-to-end (E2E) delay reduction is one important example in context of the big data applications where, the integration of big data and machine learning will assist in performing the predictive analysis for the determination of the optimized user data based paths for the reduction in E2E delays within the 6G networks \cite{chh}. 
\subsection{Backscatter Communication}
The interactions between two battery-less devices are enabled through the utility of the ambient backscatter wireless communication that is based on the available RF signals (i.e., ambient television and cellular transmissions \cite{lpt}. However, for a short communication range a reasonable data rate can be obtained and the sensor based transmission of the small monitoring signals be achieved with negligible power consumption. Due to impending connectivity associated with the battery-less nodes in the backscatter systems, its potential usability in terms of providing massive connectivity in the future 6G networks is highly implicated. Where, the acquisition of critical requirements (i.e., exact phase and channel state) at nodes within the networks cannot be neglected. Eventually, these requirements can be fulfilled through the usage of non-coherent backscatter communications that show a greater potential for the optimization of resource deployment and augmentation of the services in the network devices \cite{nsm}.

\subsection{Proactive Caching}
The most important concern for the 6G networks is the large scale deployment of small cell networks for the enhancement of the overall network properties including coverage, capacity and mobility management which will lead to the massive downlink traffic overload (at BS). Therefore, the proactive caching will overcome these limitations through the provision of reduced access delay and traffic offloading which will ultimately enhance the quality-of- user experience \cite{yhc17}. Moreover, to allow for the fruitful deployment of the 6G networks extensive research should be conducted to elucidate the joint optimization of various aspects including proactive caching, management of interference, intelligently coded schemes, and scheduling techniques that essential for the 6G networks.

\subsection{Unmanned Aerial Vehicles (UAV)}
An essential component of the 6G communication networks are the UAVs/drones. In many instances the UAVs aim to provide a very high data rate and wireless connectivity. The UAVs display a capability of providing cellular connectivity mainly due to the installation of BS entities but it is clearly evident that the certain additional features displayed by the UAVs including easy deployment, strong line of sight links, and degrees of freedom with controllable mobility are not supported by the fixed infrastructures of BS \cite{tkw}. The implementation of infrastructures based on the terrestrial communications has limited practicality and economic feasibility as it is nearly impossible to provide services during situations of emergency or natural disasters but the UAVs can manage such situations easily. Therefore, the UAVs will provide new avenues for the wireless communications as it provides facilitation for the uMUB, uHSLLC, mMTC and uHDD requirements of the wireless networks \cite{lxz18}. A broader applicability is demonstrated by UAVs that spans from strengthening network connectivity to fire detection, emergency services, disaster management, monitoring pollution/parking/accidents, security and surveillance. Owing to these facts the UAVs have been considered as one the most important technologies for the 6G networks. A UAV-enabled backscatter communication for the assistance of various communication based tasks including the supply of ambient power  and the creation of suitable channel conditions for remote sensors has been reported elsewhere \cite{zzl16}. The combined application of the non-coherent detection systems and UAVs can help in the creation of air-interfaces that are appropriate and well suited for the 6G networks. However, in order to perform the realization of UAVs for the incorporation of intelligence into the 6G networks a deep reinforcement learning based robust resource allocation procedure can be used \cite{chh}.

\subsection{Terahertz Communications}
The spectral efficiency can be enhanced by widening bandwidths and enabling applications of advanced MIMO technologies \cite{sarieddeen2020next}. The extensive applications and higher data rates is the result of 5G communications relying on mmWave frequencies. Whereas, the 6G aims to extend the frequency boundaries to THz to meet the increasing demands of the future communications. THz waves or sub-millimeter radiations usually display the frequency bands and wavelengths between 0.1 THz -10 THz and 0.03 mm–3 mm respectively \cite{h18, su19}. THz band will form an important component of the 6G communication as the RF band has now become exhausted and is nearly inadequate to meet the higher requirements of the 6G networks\cite{ajh14, tek}. For cellular communications the band range of 275 GHz–3 THz has been described as the main part of the THz band by the ITU Radio communication Sector (ITU-R)(Stoica and de Abreu 2019). The addition of THz band (275 GHz–3THz) to the existing mmWave band (30–300 GHz) would definitely increase the capacity for the 6G networks. Since the 275 GHz–3 THz band range has yet not been applied for any global functionality thus, the desired higher data rates can be potentially achieved using this band range \cite{tek}. However, the total band capacity can be increased by a minimum of 11.11x by the addition of THz band to the existing mmWave band. The 300 GHz–3 THz is a part of the optical band but it displays properties quite similar to the RF band which is mainly due to the fact that that THz lies at the boundary of the optical band that is positioned immediately after the RF band. This leads to increased potentials and challenges associated to the applicability of the THz in the 6G wireless communications \cite{gpm}. Two most critical properties of the THz are its wider applicability to achieve high data rates and a high path loss mainly due to a higher frequency \cite{mja}.
     
Additionally, the utilization of THz band will allow a fast-track and efficient provision of various services in 6G including uMUB, uHSLLC, and uHDD. That ultimately leads to the increased potential of 6G communications, through the provision of extensive support for wireless sensing, cognition, imaging, positioning, and communication procedures. The shorter THz wavelength offers the advantage of including a large number of antennas thus, offering hundreds of beams in comparison to the mmWave band \cite{chh}. The orbital angular momentum (OAM) multiplexing can be brought in to consideration to improve the overall spectral efficiency which can accomplished through the superimposition of multiple electromagnetic waves with highly diverse modes of the orbital angular momentum  \cite{sbg}.  Moreover, there also exists a possibility to reduce the co-channel aggregated interference and severe loss in propagation associated with the mmWave and THz bands through the formation of very narrow beams. The high atmospheric attenuation observed at THz based communications can be controlled significantly using the highly directional pencil beam based antennas. Hereby, the fixed aperture sized antennas deliver a squared frequency that provides an overall improvement in the gain and directionality which is definitely advantageous for the communication systems based on THz \cite{rxk, chi}.

\subsection{Optical Wireless Communications (OWC)}
Some of the eminent and well-known OWC technologies, including visible light communication (VLC), light fidelity, optical camera communication, and optical band based FSO communication, display extensive usage in several applications ( i.e., V2X communication, indoor mobile robot positioning, VR, underwater OWC) \cite{chh, chi, hcs, chh, saeed2019underwater}.  In addition to the RF-based communications, the OWCs are also intended for 6G communications, and it is also evident that these FSO  among OWC can provide network-to-backhaul/fronthaul connectivity. Due to various complexities and remote geographical locations, optical fiber-based connectivity as a backhaul network is difficult. The installation of optical fiber links for small-cell networks might not offer an economical and reasonable solution. Also, the 6G demands a huge density of users for access, to manage and control the majority of the access networks, a considerable level of integration of the backhaul and access networks is highly necessitated \cite{chs}. The utility of the FSO fronthaul/backhaul network is emerging and will be applied to the 5GB communications in the near future \cite{dda, bda, gzj}. Moreover, the FSO based system displays transmitter and receiver characteristics that are similar to that of the optical fiber networks, thus indicating that the data transfer operation in the FSO occurs in a truly self-directed and autonomous manner \cite{nsm}.

\subsection{MIMO-Cell-Free Communication}
The large intelligent surfaces (LIS) and intelligent reflecting surfaces (IRS) are two types of intelligent surfaces. Both are considered to be promising 6G candidate technologies. In \cite{8319526}, authors first presented the idea of using antenna arrays as the LIS in large MIMO systems.

Unlike beamforming, which requires a large number of antennas to focus signals, the LIS is electromagnetically proactive in the external environment and places few constraints on how antennas spread. As a result, the LIS is able to avoid the negative impacts of antenna correlations. However, because of the active property of the surfaces, the LIS consumes a lot of power and is not energy efficient.

The consolidation of different communication technologies and multiple frequencies for the 6G networks will allow the user to effortlessly shift from one network to another without the requirement of any manual configuration \cite{gpm}. For the 6G communication networks a shift from both conventional cellular and orthogonal communications would be observed towards the cell-free and non-orthogonal communications, thus, allowing for the automatic selection of the best network from the available set of communication technologies. In the current networks, the movement from one cell to another leads to various handover failures, delays and data losses which will be taken over by the 6G cell-free communications. Therefore, the utilization of multi-connectivity, multi-tier hybrid techniques and heterogeneous radio based devices will allow the effective augmentation of the Cell-free communication \cite{chh, gpm}.

\subsection{Blockchain-Security Perspective}
Blockchain is a decentralized database that are based on the hash tree theory, are tamper-proof and difficult to reverse \cite{10.1145/2994581}. Authenticity, data security, and accessibility are all characteristics of blockchain \cite{akhtar2020shift}. As a result, without the necessity of a centralised authority, blockchain can be utilised to manage spectrum resources in 6G context. Furthermore, blockchain may be used to secure data security and privacy, as well as regulate access. In \cite{fan2018blockchain}, authors offer a privacy-preserving blockchain-based approach that combines access policy and encryption technologies to ensure data privacy. In mobile cognitive radio networks \cite{kotobi2018secure}, authors use blockchain as a decentralised database to improve accessibility protocols and ensure spectral allocation.

\section{Conclusion}
%%%%%%%%%%%%%%%%%%%%%%%%%%%%%%%%%%%%%%%%%%%%%%%%%%%%%%%%%%%%%%%%%%%%%%%%%
The evolution of each generation of wireless communication networks brings enhancements to existing technologies and adds new features to meet future requirements. Although the 5G communication system displays promising features, it is still not adequate to meet the ever-growing wireless communication requirements. These factors call for envisioning the 6G networks to effectively cater to the demands of the new era of communication systems. Extensive research is carried on elucidating important aspects of the 6G networks, thus indicating a promising future utility. Therefore, we provide a brief overview of the overall 6G networks, the evolution of the communication networks, the marketing/research activities on the 6G mobile communication networks, the enabling technologies for the 6G networks, and the current state-of-the-art works on 6G communication systems. Besides providing an insight into the vision of the 6G, we have also elaborated various technologies that will form the core of 6G.  Additionally, we also focus on the emerging data rate improving technologies such as the relatively new spectrum technologies (i.e., THz communication and VLC) and new communication paradigms (i.e., molecular and quantum communication). In the face of a globally accruing digital divide, we believe that this paper can motivate researchers to investigate the enabling technologies for 6G systems and their applications in IoT.
 
%\section*{ACKNOWLEDGEMENT}

{
%\balance
\bibliographystyle{IEEEtran}
%\addcontentsline{toc}{chapter}{References}
\bibliography{IEEEabrv,Reference}

\begin{thebibliography}{100}
\providecommand{\url}[1]{#1}
\csname url@samestyle\endcsname
\providecommand{\newblock}{\relax}
\providecommand{\bibinfo}[2]{#2}
\providecommand{\BIBentrySTDinterwordspacing}{\spaceskip=0pt\relax}
\providecommand{\BIBentryALTinterwordstretchfactor}{4}
\providecommand{\BIBentryALTinterwordspacing}{\spaceskip=\fontdimen2\font plus
\BIBentryALTinterwordstretchfactor\fontdimen3\font minus
  \fontdimen4\font\relax}
\providecommand{\BIBforeignlanguage}[2]{{%
\expandafter\ifx\csname l@#1\endcsname\relax
\typeout{** WARNING: IEEEtran.bst: No hyphenation pattern has been}%
\typeout{** loaded for the language `#1'. Using the pattern for}%
\typeout{** the default language instead.}%
\else
\language=\csname l@#1\endcsname
\fi
#2}}
\providecommand{\BIBdecl}{\relax}
\BIBdecl

\bibitem{chen2020vision}
S.~Chen, Y.-C. Liang, S.~Sun, S.~Kang, W.~Cheng, and M.~Peng, ``Vision,
  requirements, and technology trend of {6G}: How to tackle the challenges of
  system coverage, capacity, user data-rate and movement speed,'' \emph{IEEE
  Wireless Communications}, vol.~27, no.~2, pp. 218--228, 2020.

\bibitem{aka}
M.~H. Alsharif, A.~H. Kelechi, M.~A. Albreem \emph{et~al.}, ``Sixth generation
  {({6G})} wireless networks: Vision, research activities, challenges and
  potential solutions. symmetry,'' \emph{Symmetry}, vol.~12, no. 676, p.~4,
  2020.

\bibitem{kyi}
L.~U. Khan, I.~Yaqoob, M.~Imran \emph{et~al.}, ``{{6G}} wireless systems: {A}
  vision, architectural elements, and future directions,'' \emph{IEEE Access},
  vol.~8, no.~14, pp. 47\,029--14\,704, 2020.

\bibitem{9380673}
G.~Karabulut~Kurt, M.~G. Khoshkholgh, S.~Alfattani, A.~Ibrahim, T.~S.~J.
  Darwish, M.~S. Alam, H.~Yanikomeroglu, and A.~Yongacoglu, ``A vision and
  framework for the high altitude platform station (haps) networks of the
  future,'' \emph{IEEE Communications Surveys Tutorials}, vol.~23, no.~2, pp.
  729--779, 2021.

\bibitem{9530717}
S.~Basharat, S.~Ali~Hassan, H.~Pervaiz, A.~Mahmood, Z.~Ding, and M.~Gidlund,
  ``Reconfigurable intelligent surfaces: Potentials, applications, and
  challenges for {6G} wireless networks,'' \emph{IEEE Wireless Communications},
  pp. 1--8, 2021.

\bibitem{wwc}
H.~Wang, W.~Wang, X.~Chen \emph{et~al.}, \emph{Wireless information and energy
  transfer in interference aware massive MIMO systems.}\hskip 1em plus 0.5em
  minus 0.4em\relax IEEE Global Communications Conference, 2014.

\bibitem{ywh}
X.~You, C.-X. Wang, J.~Huang \emph{et~al.}, ``Towards {{6G}} wireless
  communication networks: Vision, enabling technologies, and new paradigm
  shifts.'' \emph{Science China Information Sciences}, vol.~64, no.~1, pp.
  1--74, 2021.

\bibitem{9524814}
V.-L. Nguyen, P.-C. Lin, B.-C. Cheng, R.-H. Hwang, and Y.-D. Lin, ``Security
  and privacy for {{6G}}: A survey on prospective technologies and
  challenges,'' \emph{IEEE Communications Surveys Tutorials}, pp. 1--1, 2021.

\bibitem{9509294}
D.~C. Nguyen, M.~Ding, P.~N. Pathirana, A.~Seneviratne, J.~Li, D.~Niyato,
  O.~Dobre, and H.~V. Poor, ``{{6G}} internet of things: A comprehensive
  survey,'' \emph{IEEE Internet of Things Journal}, pp. 1--1, 2021.

\bibitem{9358097}
N.-N. Dao, Q.-V. Pham, N.~H. Tu, T.~T. Thanh, V.~N.~Q. Bao, D.~S. Lakew, and
  S.~Cho, ``Survey on aerial radio access networks: Toward a comprehensive
  {{6G}} access infrastructure,'' \emph{IEEE Communications Surveys Tutorials},
  vol.~23, no.~2, pp. 1193--1225, 2021.

\bibitem{9397776}
C.~D. Alwis, A.~Kalla, Q.-V. Pham, P.~Kumar, K.~Dev, W.-J. Hwang, and
  M.~Liyanage, ``Survey on {6G} frontiers: Trends, applications, requirements,
  technologies and future research,'' \emph{IEEE Open Journal of the
  Communications Society}, vol.~2, pp. 836--886, 2021.

\bibitem{9403380}
F.~Tang, B.~Mao, Y.~Kawamoto, and N.~Kato, ``Survey on machine learning for
  intelligent end-to-end communication toward {6G}: From network access,
  routing to traffic control and streaming adaption,'' \emph{IEEE
  Communications Surveys Tutorials}, vol.~23, no.~3, pp. 1578--1598, 2021.

\bibitem{9369324}
F.~Guo, F.~R. Yu, H.~Zhang, X.~Li, H.~Ji, and V.~C.~M. Leung, ``Enabling
  massive iot toward {6G}: A comprehensive survey,'' \emph{IEEE Internet of
  Things Journal}, vol.~8, no.~15, pp. 11\,891--11\,915, 2021.

\bibitem{9385374}
X.~Fang, W.~Feng, T.~Wei, Y.~Chen, N.~Ge, and C.-X. Wang, ``5g embraces
  satellites for {6G} ubiquitous iot: Basic models for integrated satellite
  terrestrial networks,'' \emph{IEEE Internet of Things Journal}, vol.~8,
  no.~18, pp. 14\,399--14\,417, 2021.

\bibitem{9184022}
S.~Aggarwal, N.~Kumar, and S.~Tanwar, ``Blockchain-envisioned uav communication
  using {6G} networks: Open issues, use cases, and future directions,''
  \emph{IEEE Internet of Things Journal}, vol.~8, no.~7, pp. 5416--5441, 2021.

\bibitem{9200376}
A.~H. Sodhro, S.~Pirbhulal, Z.~Luo, K.~Muhammad, and N.~Z. Zahid, ``Toward {6G}
  architecture for energy-efficient communication in iot-enabled smart
  automation systems,'' \emph{IEEE Internet of Things Journal}, vol.~8, no.~7,
  pp. 5141--5148, 2021.

\bibitem{9355403}
Y.~Xu, G.~Gui, H.~Gacanin, and F.~Adachi, ``A survey on resource allocation for
  5g heterogeneous networks: Current research, future trends, and challenges,''
  \emph{IEEE Communications Surveys Tutorials}, vol.~23, no.~2, pp. 668--695,
  2021.

\bibitem{9328851}
J.~Liang, L.~Li, and C.~Zhao, ``A transfer learning approach for compressed
  sensing in {6G}-iot,'' \emph{IEEE Internet of Things Journal}, pp. 1--1,
  2021.

\bibitem{9044345}
B.~Mao, Y.~Kawamoto, and N.~Kato, ``Ai-based joint optimization of qos and
  security for {6G} energy harvesting internet of things,'' \emph{IEEE Internet
  of Things Journal}, vol.~7, no.~8, pp. 7032--7042, 2020.

\bibitem{hzl}
X.~Huang, J.~A. Zhang, R.~P. Liu \emph{et~al.}, ``Integrating space and
  terrestrial networks with passenger airplanes for the generation
  wireless-will it work?'' \emph{IEEE Vehicular Technology Magazine}, vol.~6,
  2019.

\bibitem{ars16}
M.~Agiwal, A.~Roy, and N.~Saxena, ``Next generation {5G} wireless networks: {A}
  comprehensive survey,'' \emph{IEEE Communications Surveys and Tutorials},
  vol.~18, no.~3, pp. 1617--1655, 2016.

\bibitem{lxz18}
S.~Li, L.~D. Xu, and S.~Zhao, ``{5G} internet of things: {A} survey,''
  \emph{Journal of Industrial Information Integration}, vol.~10, pp. 1--9,
  2018.

\bibitem{sai}
S.~A.~A. Shah, E.~Ahmed, M.~Imran \emph{et~al.}, ``5g for vehicular
  communications,'' \emph{IEEE communications magazine}, vol.~56, no.~1, pp.
  111--117, 2018.

\bibitem{pb11}
J.~Parikh and A.~Basu, ``Lte advanced: The 4g mobile broadband technology,''
  \emph{International Journal of Computer Applications}, vol.~13, no.~5, pp.
  17--21, 2011.

\bibitem{sms}
M.~Shafi, A.~F. Molisch, P.~J. Smith \emph{et~al.}, ``5g: A tutorial overview
  of standards, trials, challenges, deployment, and practice,'' \emph{IEEE
  journal on selected areas in communications}, vol.~35, no.~6, pp. 1201--1221,
  2017.

\bibitem{wdo}
J.~Wu, M.~Dong, K.~Ota \emph{et~al.}, ``Big data analysis-based secure cluster
  management for optimized control plane in software-defined networks,''
  \emph{IEEE Transactions on Network and Service Management}, vol.~15, no.~1,
  pp. 27--38, 2018.

\bibitem{ykk18}
A.~Yastrebova, R.~Kirichek, Y.~Koucheryavy \emph{et~al.}, ``Future networks
  2030: Architecture and requirements.'' \emph{10th International Congress on
  Ultra Modern Telecommunications and Control Systems and Workshops (ICUMT),
  IEEE.}, 2018.

\bibitem{zfw}
B.~Zong, C.~Fan, X.~Wang \emph{et~al.}, ``{{6G}} technologies: Key drivers,
  core requirements, system architectures, and enabling technologies,''
  \emph{IEEE Vehicular Technology Magazine}, vol.~14, no.~3, pp. 18--27, 2019.

\bibitem{r20}
S.~P. Rout, ``{{6G}} wireless communication: Its vision, viability,
  application, requirement, technologies, encounters and research.'' in
  \emph{Viability}, 11th International Conference on Computing, Communication
  and Networking Technologies (ICCCNT), IEEE, 2020.

\bibitem{jit}
M.~Jaber, M.~A. Imran, R.~Tafazolli \emph{et~al.}, ``{5G} backhaul challenges
  and emerging research directions: {A} survey,'' \emph{IEEE Access}, vol.~4,
  pp. 1743--1766, 2016.

\bibitem{chs}
M.~Z. Chowdhury, M.~K. Hasan, M.~Shahjalal \emph{et~al.}, ``Optical wireless
  hybrid networks: Trends, opportunities, challenges, and research
  directions,'' \emph{IEEE Communications Surveys and Tutorials}, vol.~22,
  no.~2, pp. 930--966, 2020.

\bibitem{kyt}
L.~U. Khan, I.~Yaqoob, N.~H. Tran \emph{et~al.}, ``Edge-computing-enabled smart
  cities: {A} comprehensive survey,'' \emph{IEEE Internet of Things Journal},
  vol.~7, no.~10, pp. 10\,200--10\,232, 2020.

\bibitem{krz}
W.~Z. Khan, M.~Rehman, H.~M. Zangoti \emph{et~al.}, ``Industrial internet of
  things: Recent advances, enabling technologies and open challenges,''
  \emph{Computers and Electrical Engineering}, vol.~81, p. 10652.

\bibitem{lz20}
Y.~Lu and X.~Zheng, ``{{6G}}: {A} survey on technologies, scenarios,
  challenges, and the related issues,'' \emph{Journal of Industrial Information
  Integration}, vol. 100158, 2020.

\bibitem{ln20}
Y.~Lu and X.~Ning, ``A vision of {{6G}-5G's} successor,'' \emph{Journal of
  Management Analytics}, vol.~7, no.~3, pp. 301--320, 2020.

\bibitem{zxm}
Z.~Zhang, Y.~Xiao, Z.~Ma \emph{et~al.}, ``{{6G}} wireless networks: Vision,
  requirements, architecture, and key technologies,'' \emph{IEEE Vehicular
  Technology Magazine}, vol.~14, no.~3, pp. 28--41, 2019.

\bibitem{hl20}
M.~Hensmans and G.~Liu, ``Huawei's long march to global leadership: Joint
  innovation strategy from the periphery to the center,'' in \emph{Huawei Goes
  Global}.\hskip 1em plus 0.5em minus 0.4em\relax springer, 2020, pp. 225--245.

\bibitem{akn20}
I.~F. Akyildiz, A.~Kak, and S.~Nie, ``{{6G}} and beyond: {The} future of
  wireless communications systems,'' \emph{IEEE Access}, vol.~8, no.~10, pp.
  33\,995--13\,403, 2020.

\bibitem{h20}
J.~Hayes, ``Network-communication {{6G}} and the reinvention of mobile,''
  \emph{Engineering and Technology}, vol.~15, no.~1, pp. 26--29, 2020.

\bibitem{sbc19}
W.~Saad, M.~Bennis, and M.~Chen, ``A vision of {{6G}} wireless systems:
  Applications, trends, technologies, and open research problems,'' \emph{IEEE
  network}, vol.~34, no.~3, pp. 134--142, 2019.

\bibitem{lcs}
K.~B. Letaief, W.~Chen, Y.~Shi \emph{et~al.}, ``The roadmap to {{6G}}: Ai
  empowered wireless networks.'' \emph{IEEE communications magazine}, vol.~57,
  no.~8, pp. 84--90, 2019.

\bibitem{tkw}
F.~Tariq, M.~R. Khandaker, K.-K. Wong \emph{et~al.}, ``A speculative study on
  {{6G}}.'' \emph{IEEE Wireless Communications}, vol.~27, no.~4, pp. 118--125.

\bibitem{gpm}
M.~Giordani, M.~Polese, M.~Mezzavilla \emph{et~al.}, ``Toward {{6G}} networks:
  Use cases and technologies.'' \emph{IEEE communications magazine}, vol.~58,
  no.~3, pp. 55--61, 2020.

\bibitem{yxx}
P.~Yang, Y.~Xiao, M.~Xiao \emph{et~al.}, ``{{6G}} wireless communications:
  Vision and potential techniques,'' \emph{IEEE network}, vol.~33, no.~4, pp.
  70--75, 2019.

\bibitem{zln19}
L.~Zhang, Y.-C. Liang, and D.~Niyato, ``{{6G}} visions: Mobile ultra-broadband,
  super internet-of-things, and artificial intelligence,'' \emph{China
  Communications}, vol.~16, no.~8, pp. 1--14, 2019.

\bibitem{kmt}
N.~Kato, B.~Mao, F.~Tang \emph{et~al.}, ``Ten challenges in advancing machine
  learning technologies toward {{6G}}.'' \emph{IEEE Wireless Communications},
  vol.~27, no.~3, pp. 96--103, 2020.

\bibitem{ajh14}
I.~F. Akyildiz, J.~M. Jornet, and C.~Han, ``Terahertz band: Next frontier for
  wireless communications,'' \emph{Physical Communication}, vol.~12, pp.
  16--32, 2014.

\bibitem{cls}
S.~Chen, Y.-C. Liang, S.~Sun \emph{et~al.}, ``Vision, requirements, and
  technology trend of {{6G}}: How to tackle the challenges of system coverage,
  capacity, user data-rate and movement speed,'' in \emph{Vision}.\hskip 1em
  plus 0.5em minus 0.4em\relax IEEE Wireless Communications 27(2), pp.
  218--228.

\bibitem{nsm}
S.~J. Nawaz, S.~K. Sharma, B.~Mansoor \emph{et~al.}, ``Non-coherent and
  backscatter communications: Enabling ultra-massive connectivity in {{6G}}
  wireless networks.'' \emph{IEEE Access}, vol.~6.

\bibitem{sbg}
E.~C. Strinati, S.~Barbarossa, J.~L. Gonzalez-Jimenez \emph{et~al.}, ``{{6G}}:
  The next frontier: From holographic messaging to artificial intelligence
  using subterahertz and visible light communication,'' \emph{IEEE Vehicular
  Technology Magazine}, vol.~14, no.~3, pp. 42--50, 2019.

\bibitem{sh19}
M.~Salehi and E.~Hossain, ``On the effect of temporal correlation on joint
  success probability and distribution of number of interferers in mobile uav
  networks,'' \emph{IEEE Wireless Communications Letters}, vol.~8, no.~6, pp.
  1621--1625, 2019.

\bibitem{hyw}
T.~Huang, W.~Yang, J.~Wu \emph{et~al.}, ``A survey on green {{6G}} network:
  Architecture and technologies.'' \emph{IEEE Access}, vol.~7, no.~18, pp.
  75\,758--17\,576, 2019.

\bibitem{ekm19}
D.~Elliott, W.~Keen, and L.~Miao, ``Recent advances in connected and automated
  vehicles.'' \emph{journal of traffic and transportation engineering}, vol.~6,
  no.~2, pp. 109--131, 2019.

\bibitem{jlz}
B.~Ji, Y.~Li, B.~Zhou \emph{et~al.}, ``Performance analysis of uav relay
  assisted iot communication network enhanced with energy harvesting,''
  \emph{IEEE Access}, vol.~7, pp. 38\,738--38\,747, 2019.

\bibitem{chh}
M.~Z. Chowdhury, M.~T. Hossan, M.~K. Hasan \emph{et~al.}, ``Integrated
  rf/optical wireless networks for improving qos in indoor and transportation
  applications,'' \emph{Wireless Personal Communications}, vol. 107, no.~3, pp.
  1401--1430, 2019.

\bibitem{llp}
L.~Lov{\'e}n, T.~Lepp{"a}nen, E.~Peltonen \emph{et~al.}, ``Edgeai: A vision for
  distributed, edgenative artificial intelligence in future {{6G}} networks.''
  \emph{The 1st {6G} Wireless Summit}, vol.~6, pp. 1--2, 2019.

\bibitem{cml}
F.~Clazzer, A.~Munari, G.~Liva \emph{et~al.}, ``From {5G} to {{6G}}: Has the
  time for modern random access come?'' Tech. Rep., 2019.

\bibitem{vm20}
H.~Viswanathan and P.~E. Mogensen, ``Communications in the {{6G}} era,''
  \emph{IEEE Access}, vol.~8, pp. 57\,063--57\,074, 2020.

\bibitem{mal20}
N.~H. Mahmood, H.~Alves, O.~A. L{\'o}pez \emph{et~al.}, ``Six key features of
  machine type communication in {6G}.x,'' \emph{G2nd {{6G}} Wireless Summit
  ({6G} SUMMIT), IEEE}, vol.~6, 2020.

\bibitem{das}
S.~Dang, O.~Amin, B.~Shihada \emph{et~al.}, ``From a human-centric perspective:
  What might {{6G}} be?''

\bibitem{zll}
H.~Zhang, Y.~Li, Z.~Lv \emph{et~al.}, ``A real-time and ubiquitous network
  attack detection based on deep belief network and support vector machine,''
  \emph{IEEE/CAA Journal of Automatica Sinica}, vol.~7, no.~3, pp. 790--799.

\bibitem{zdw}
Y.~Zhang, B.~Di, P.~Wang \emph{et~al.}, ``Hetmec: Heterogeneous multi-layer
  mobile edge computing in the {{6G}} era,'' \emph{IEEE Transactions on
  Vehicular Technology}, vol.~69, no.~4, pp. 4388--4400.

\bibitem{glt}
G.~Gui, M.~Liu, F.~Tang \emph{et~al.}, ``{{6G}}: Opening new horizons for
  integration of comfort, security, and intelligence,'' \emph{IEEE Wireless
  Communications}, vol.~27, no.~5, pp. 126--132, 2020.

\bibitem{tkp}
I.~Tomkos, D.~Klonidis, E.~Pikasis \emph{et~al.}, ``Toward the {{6G}} network
  era: Opportunities and challenges.'' \emph{IT Professional}, vol.~22, no.~1,
  pp. 34--38, 2020.

\bibitem{ya20}
E.~Yaacoub and M.-S. Alouini, ``A key {{6G}} challenge and
  opportunity---connecting the base of the pyramid: A survey on rural
  connectivity,'' in \emph{Proceedings of the IEEE 108(4}, 2020, pp. 533--582.

\bibitem{slc}
R.~Shafin, L.~Liu, V.~Chandrasekhar \emph{et~al.}, ``Artificial
  intelligence-enabled cellular networks: A critical path to beyond-5g and
  {6G}.'' \emph{IEEE Wireless Communications}, vol.~27, no.~2, pp. 212--217.

\bibitem{csa}
M.~Z. Chowdhury, M.~Shahjalal, S.~Ahmed \emph{et~al.}, ``{{6G}} wireless
  communication systems: Applications, requirements, technologies, challenges,
  and research directions,'' \emph{IEEE Open Journal of the Communications
  Society}, vol.~1, pp. 957--975, 2020.

\bibitem{k21}
J.~H. Kim, ``{6G} and internet of things: a survey,'' \emph{Journal of
  Management Analytics}, pp. 1--17, 2021.

\bibitem{aj21}
Z.~Allam and D.~S. Jones, ``Future (post-covid) digital, smart and sustainable
  cities in the wake of {{6G}}: Digital twins, immersive realities and new
  urban economies,'' \emph{Land Use Policy}, vol. 101, no. 10520, p.~1, 2021.

\bibitem{pc21}
P.~K. Padhi and F.~Charrua-Santos, ``{{6G}} enabled industrial internet of
  everything: Towards a theoretical framework,'' \emph{Applied System
  Innovation}, vol.~4, no.~1, p.~11, 2021.

\bibitem{ycw}
Z.~Yang, M.~Chen, K.-K. Wong \emph{et~al.}, ``Federated learning for {{6G}}:
  Applications, challenges, and opportunities.'' Tech. Rep., 2021.

\bibitem{wth}
Y.~Wang, Y.~Tian, X.~Hei \emph{et~al.}, ``A novel iov block-streaming service
  awareness and trusted verification in {{6G}}.'' \emph{IEEE Transactions on
  Vehicular Technology}, vol.~6, 2021.

\bibitem{sap}
A.~Shahraki, M.~Abbasi, M.~Piran \emph{et~al.}, ``A comprehensive survey on
  {{6G}} networks: Applications, core services, enabling technologies, and
  future challenges. arxiv preprint,'' Tech. Rep., 2021.

\bibitem{iat}
A.~L. Imoize, O.~Adedeji, N.~Tandiya \emph{et~al.}, ``{{6G}} enabled smart
  infrastructure for sustainable society: Opportunities, challenges, and
  research roadmap,'' \emph{Sensors}, vol.~21, no.~5, p. 1709, 2021.

\bibitem{w21}
H.~Wang, ``Application of data mining technology in quality evaluation of
  online teaching based on {{6G}},'' \emph{The Educational Review, USA},
  vol.~5, no.~2, pp. 27--30, 2021.

\bibitem{db18}
K.~David and H.~Berndt, ``{{6G}} vision and requirements: Is there any need for
  beyond 5g?'' \emph{IEEE Vehicular Technology Magazine}, vol.~13, no.~3, pp.
  72--80, 2018.

\bibitem{hcz18}
F.~Hu, B.~Chen, and K.~Zhu, ``Full spectrum sharing in cognitive radio networks
  toward {5G}: {A} survey,'' \emph{IEEE Access}, vol.~6, pp. 15\,754--15\,776,
  2018.

\bibitem{lfz18}
B.~Li, Z.~Fei, and Y.~Zhang, ``Uav communications for {5G} and beyond: Recent
  advances and future trends,'' \emph{IEEE Internet of Things Journal}, vol.~6,
  no.~2, pp. 2241--2263, 2018.

\bibitem{y15}
H.~You, \emph{Key parameters for 5G mobile communications [ITU-R WP 5D
  standardization status]}.\hskip 1em plus 0.5em minus 0.4em\relax Seongnam-Si,
  South Korea, Tech. Rep: KT. Korea Telecom, 2015.

\bibitem{np21}
S.~Nayak and R.~Patgiri, ``{{6G}} communication technology: A vision on
  intelligent healthcare,'' in \emph{Health Informatics: A Computational
  Perspective in Healthcare}, 2021, pp. 1--18.

\bibitem{kgt21}
A.~Kumari, R.~Gupta, and S.~Tanwar, \emph{Amalgamation of blockchain and IoT
  for smart cities underlying {{6G}} communication: A comprehensive
  review}.\hskip 1em plus 0.5em minus 0.4em\relax Computer Communications,
  2021.

\bibitem{asr}
S.~Ali, W.~Saad, N.~Rajatheva \emph{et~al.}, ``{{6G}} white paper on machine
  learning in wireless communication networks. arxiv,'' 2020, preprint.

\bibitem{saeed2020cubesat}
N.~Saeed, A.~Elzanaty, H.~Almorad, H.~Dahrouj, T.~Y. Al-Naffouri, and M.-S.
  Alouini, ``Cubesat communications: {R}ecent advances and future challenges,''
  \emph{IEEE Communications Surveys and Tutorials}, vol.~22, no.~3, pp.
  1839--1862, 2020.

\bibitem{8907365}
M.~A. Al-Jarrah, M.~A. Yaseen, A.~Al-Dweik, O.~A. Dobre, and E.~Alsusa,
  ``Decision fusion for iot-based wireless sensor networks,'' \emph{IEEE
  Internet of Things Journal}, vol.~7, no.~2, pp. 1313--1326, 2020.

\bibitem{9512099}
``“internet available: of things 2016,” 2016. [online].
  https://www.cisco.com/c/dam/en/us/products/collateral/se/
  internetof-things/at-a-glance-c45-731471.pdf.'' \emph{Cisco}.

\bibitem{9512098}
``“number to 125 of connected billion iot by devices 2030,” will 2021.
  surge [online]. available:
  https://news.ihsmarkit.com/prviewer/releaseonly/slug/
  number-connected-iot-devices-will-surge-125-billion-2030.''

\bibitem{7397856}
M.~R. Palattella, M.~Dohler, A.~Grieco, G.~Rizzo, J.~Torsner, T.~Engel, and
  L.~Ladid, ``Internet of things in the 5g era: Enablers, architecture, and
  business models,'' \emph{IEEE Journal on Selected Areas in Communications},
  vol.~34, no.~3, pp. 510--527, 2016.

\bibitem{9524496}
A.~Jagannath, J.~Jagannath, and T.~Melodia, ``Redefining wireless communication
  for {6G}: Signal processing meets deep learning with deep unfolding,''
  \emph{IEEE Transactions on Artificial Intelligence}, pp. 1--1, 2021.

\bibitem{sa19}
R.-A. Stoica and G.~T.~F. de~Abreu, ``{{6G}}: the wireless communications
  network for collaborative and ai applications.'' 2019, preprint.

\bibitem{z19}
J.~Zhao, ``A survey of intelligent reflecting surfaces (irss): Towards {{6G}}
  wireless communication networks. arxiv,'' 2019, preprint.

\bibitem{mal}
N.~H. Mahmood, H.~Alves, O.~A. L{\'o}pez \emph{et~al.}, ``Six key enablers for
  machine type communication in {{6G}}.'' Tech. Rep., 2019.

\bibitem{whg}
C.-X. Wang, F.~Haider, X.~Gao \emph{et~al.}, ``Cellular architecture and key
  technologies for 5g wireless communication networks.'' \emph{IEEE
  communications magazine}, vol.~52, no.~2, pp. 122--130, 2014.

\bibitem{jkc15}
T.~Jung, T.~Kwon, and C.-B. Chae, ``Qoe-based transmission strategies for
  multi-user wireless information and power transfer,'' \emph{ICT Express},
  vol.~1, no.~3, pp. 116--120, 2015.

\bibitem{es20}
S.~Elmeadawy and R.~M. Shubair, ``Enabling technologies for {{6G}} future
  wireless communications: Opportunities and challenges. arxiv,'' 2020,
  preprint.

\bibitem{mba20}
A.~Mitra, S.~Biswas, T.~Adhikari \emph{et~al.}, ``Emergence of edge computing:
  An advancement over cloud and fog11th international conference on
  computing.''\hskip 1em plus 0.5em minus 0.4em\relax IEEE: Communication and
  Networking Technologies (ICCCNT), 2020.

\bibitem{gei}
S.~George, T.~Eiszler, R.~Iyengar \emph{et~al.}, ``Openrtist: End-to-end
  benchmarking for edge computing,'' \emph{IEEE Pervasive Computing}, vol.~19,
  no.~4, pp. 10--18, 2020.

\bibitem{llt}
S.~Liu, L.~Liu, J.~Tang \emph{et~al.}, ``Edge computing for autonomous driving:
  Opportunities and challenges,'' \emph{Proceedings of the IEEE}, vol. 107,
  no.~8, pp. 1697--1716, 2019.

\bibitem{kad18}
N.~Kalatzis, M.~Avgeris, D.~Dechouniotis \emph{et~al.}, ``Edge computing in iot
  ecosystems for uav-enabled early fire detection.''\hskip 1em plus 0.5em minus
  0.4em\relax IEEE: IEEE International Conference on Smart Computing
  (SMARTCOMP),, 2018.

\bibitem{nae18}
G.~B. Neumann, V.~P.~D. Almeida, and M.~Endler, \emph{Smart Forests: fire
  detection service. 2018 IEEE symposium on computers and communications
  (ISCC)}.\hskip 1em plus 0.5em minus 0.4em\relax IEEE, 2018.

\bibitem{blk}
H.~Bura, N.~Lin, N.~Kumar \emph{et~al.}, \emph{An edge based smart parking
  solution using camera networks and deep learning.}\hskip 1em plus 0.5em minus
  0.4em\relax IEEE International Conference on Cognitive Computing (ICCC),
  2018.

\bibitem{kzp}
R.~Ke, Y.~Zhuang, Z.~Pu \emph{et~al.}, \emph{A smart, efficient, and reliable
  parking surveillance system with edge artificial intelligence on IoT
  devices.}\hskip 1em plus 0.5em minus 0.4em\relax IEEE Transactions on
  Intelligent Transportation Systems., 2020.

\bibitem{clx18}
X.~Chang, W.~Li, C.~Xia \emph{et~al.}, ``From insight to impact: Building a
  sustainable edge computing platform for smart homes.''\hskip 1em plus 0.5em
  minus 0.4em\relax IEEE 24th International Conference on Parallel and
  Distributed Systems (ICPADS), 2018.

\bibitem{cd17}
T.~Chakraborty and S.~K. Datta, \emph{Home automation using edge computing and
  internet of things.}\hskip 1em plus 0.5em minus 0.4em\relax IEEE
  International Symposium on Consumer Electronics (ISCE), 2017.

\bibitem{scz}
W.~Shi, J.~Cao, Q.~Zhang \emph{et~al.}, ``Edge computing: Vision and
  challenges,'' \emph{IEEE Internet of Things Journal}, vol.~3, no.~5, pp.
  637--646, 2016.

\bibitem{kck18}
M.~Kobayashi, G.~Caire, and G.~Kramer, \emph{Joint state sensing and
  communication: Optimal tradeoff for a memoryless case.}\hskip 1em plus 0.5em
  minus 0.4em\relax IEEE International Symposium on Information Theory (ISIT),
  2018.

\bibitem{sgw}
X.~Shen, J.~Gao, W.~Wu \emph{et~al.}, ``Ai-assisted network-slicing based
  next-generation wireless networks,'' \emph{IEEE Open Journal of Vehicular
  Technology}, vol.~1, pp. 45--66.

\bibitem{b17}
E.~J. Black, ``Holographic beam forming and mimo,'' in \emph{Pivotal
  Commware}.\hskip 1em plus 0.5em minus 0.4em\relax unpublished, 2017.

\bibitem{lpt}
V.~Liu, A.~Parks, V.~Talla \emph{et~al.}, ``Ambient backscatter: Wireless
  communication out of thin air,'' \emph{ACM SIGCOMM Computer Communication
  Review}, vol.~43, no.~4, pp. 39--50, 2013.

\bibitem{yhc17}
C.~Yi, S.~Huang, and J.~Cai, ``An incentive mechanism integrating joint power,
  channel and link management for social-aware d2d content sharing and
  proactive caching,'' \emph{IEEE Transactions on Mobile Computing}, vol.~17,
  no.~4, pp. 789--802, 2017.

\bibitem{zzl16}
Y.~Zeng, R.~Zhang, and T.~J. Lim, ``Wireless communications with unmanned
  aerial vehicles: Opportunities and challenges,'' \emph{IEEE communications
  magazine}, vol.~54, no.~5, pp. 36--42, 2016.

\bibitem{sarieddeen2020next}
H.~Sarieddeen, N.~Saeed, T.~Y. Al-Naffouri, and M.-S. Alouini, ``Next
  generation terahertz communications: {A} rendezvous of sensing, imaging, and
  localization,'' \emph{IEEE Communications Magazine}, vol.~58, no.~5, pp.
  69--75, 2020.

\bibitem{h18}
S.~Hanna, ``Technological and regulatory developments for electromagnetic
  transmission into the millimeter wave and terahertz wave spectrum,'' in
  \emph{Proceedings of the Future Technologies Conference}, 2018.

\bibitem{su19}
I.~Siaud and A.-M. Ulmer-Moll, \emph{THz Communications: an overview and
  challenges.}, 2019.

\bibitem{tek}
K.~Tekb{\i}y{\i}k, A.~R. Ekti, G.~K. Kurt \emph{et~al.}, ``Terahertz band
  communication systems: Challenges, novelties and standardization efforts,''
  \emph{Physical Communication}, vol.~35, p. 10070, 2019.

\bibitem{mja}
S.~Mumtaz, J.~M. Jornet, J.~Aulin \emph{et~al.}, ``Terahertz communication for
  vehicular networks,'' \emph{IEEE Transactions on Vehicular Technology},
  vol.~66, p.~7.

\bibitem{rxk}
T.~S. Rappaport, Y.~Xing, O.~Kanhere \emph{et~al.}, ``Wireless communications
  and applications above ghz: Opportunities and challenges for {{6G}} and
  beyond.'' \emph{IEEE Access}, vol. 100, pp. 78\,729--78\,757, 2019.

\bibitem{chi}
M.~Z. Chowdhury, M.~T. Hossan, A.~Islam \emph{et~al.}, ``A comparative survey
  of optical wireless technologies: Architectures and applications,''
  \emph{IEEE Access}, vol.~6, pp. 9819--9840, 2018.

\bibitem{hcs}
M.~T. Hossan, M.~Z. Chowdhury, M.~Shahjalal \emph{et~al.}, ``Human bond
  communication with head-mounted displays: Scope, challenges, solutions, and
  applications,'' \emph{IEEE communications magazine}, vol.~57, no.~2, pp.
  26--32, 2019.

\bibitem{saeed2019underwater}
N.~Saeed, A.~Celik, T.~Y. Al-Naffouri, and M.-S. Alouini, ``Underwater optical
  wireless communications, networking, and localization: A survey,'' \emph{Ad
  Hoc Networks}, vol.~94, p. 101935, 2019.

\bibitem{dda}
A.~Douik, H.~Dahrouj, T.~Y. Al-Naffouri \emph{et~al.}, ``Hybrid
  radio/free-space optical design for next generation backhaul systems,''
  \emph{IEEE Transactions on Communications}, vol.~64, no.~6, pp. 2563--2577,
  2016.

\bibitem{bda}
B.~Bag, A.~Das, I.~S. Ansari \emph{et~al.}, ``Performance analysis of hybrid
  fso systems using fso/rf-fso link adaptation,'' \emph{IEEE Photonics
  Journal}, vol.~10, no.~3, pp. 1--17, 2018.

\bibitem{gzj}
Z.~Gu, J.~Zhang, Y.~Ji \emph{et~al.}, ``Network topology reconfiguration for
  fso-based fronthaul/backhaul in g+ wireless networks.'' \emph{IEEE Access},
  vol.~6, pp. 69\,426--69\,437.

\bibitem{8319526}
S.~Hu, F.~Rusek, and O.~Edfors, ``Beyond massive mimo: The potential of data
  transmission with large intelligent surfaces,'' \emph{IEEE Transactions on
  Signal Processing}, vol.~66, no.~10, pp. 2746--2758, 2018.

\bibitem{10.1145/2994581}
\BIBentryALTinterwordspacing
S.~Underwood, ``Blockchain beyond bitcoin,'' \emph{Commun. ACM}, vol.~59,
  no.~11, p. 15–17, Oct. 2016. [Online]. Available:
  \url{https://doi.org/10.1145/2994581}
\BIBentrySTDinterwordspacing

\bibitem{akhtar2020shift}
M.~W. Akhtar, S.~A. Hassan, R.~Ghaffar, H.~Jung, S.~Garg, and M.~S. Hossain,
  ``The shift to {6G} communications: vision and requirements,''
  \emph{Human-centric Computing and Information Sciences}, vol.~10, no.~1, pp.
  1--27, 2020.

\bibitem{fan2018blockchain}
K.~Fan, Y.~Ren, Y.~Wang, H.~Li, and Y.~Yang, ``Blockchain-based efficient
  privacy preserving and data sharing scheme of content-centric network in
  5g,'' \emph{IET communications}, vol.~12, no.~5, pp. 527--532, 2018.

\bibitem{kotobi2018secure}
K.~Kotobi and S.~G. Bilen, ``Secure blockchains for dynamic spectrum access: A
  decentralized database in moving cognitive radio networks enhances security
  and user access,'' \emph{ieee vehicular technology magazine}, vol.~13, no.~1,
  pp. 32--39, 2018.

\end{thebibliography}
}

\vskip -2\baselineskip plus -1fil
\begin{IEEEbiography}[{\vspace{1 mm} \includegraphics[width=1.5in,
height=1.3in,clip, keepaspectratio]{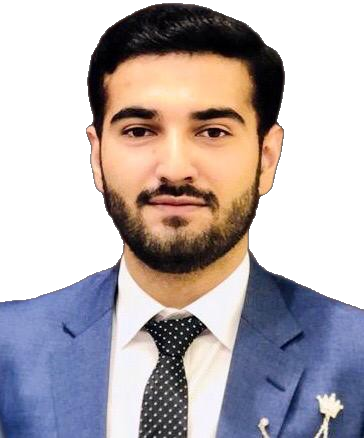}}]{Zakria Qadir} received his M.Sc. degree in Sustainable Environment and Energy Systems from Middle East Technical University, Turkey in 2019. He is currently pursuing his Ph.D. degree in wireless communication and cloud computing from the Western Sydney University, Australia. His research interests include sustainable cities, artificial intelligence, machine learning, optimization techniques, wireless communication, Internet of things, renewable energy technology, and cloud computing.
\end{IEEEbiography}

\vskip -2\baselineskip plus -1fil
\begin{IEEEbiography}[{\vspace{1 mm} \includegraphics[width=1.5in,
height=1.3in,clip, keepaspectratio]{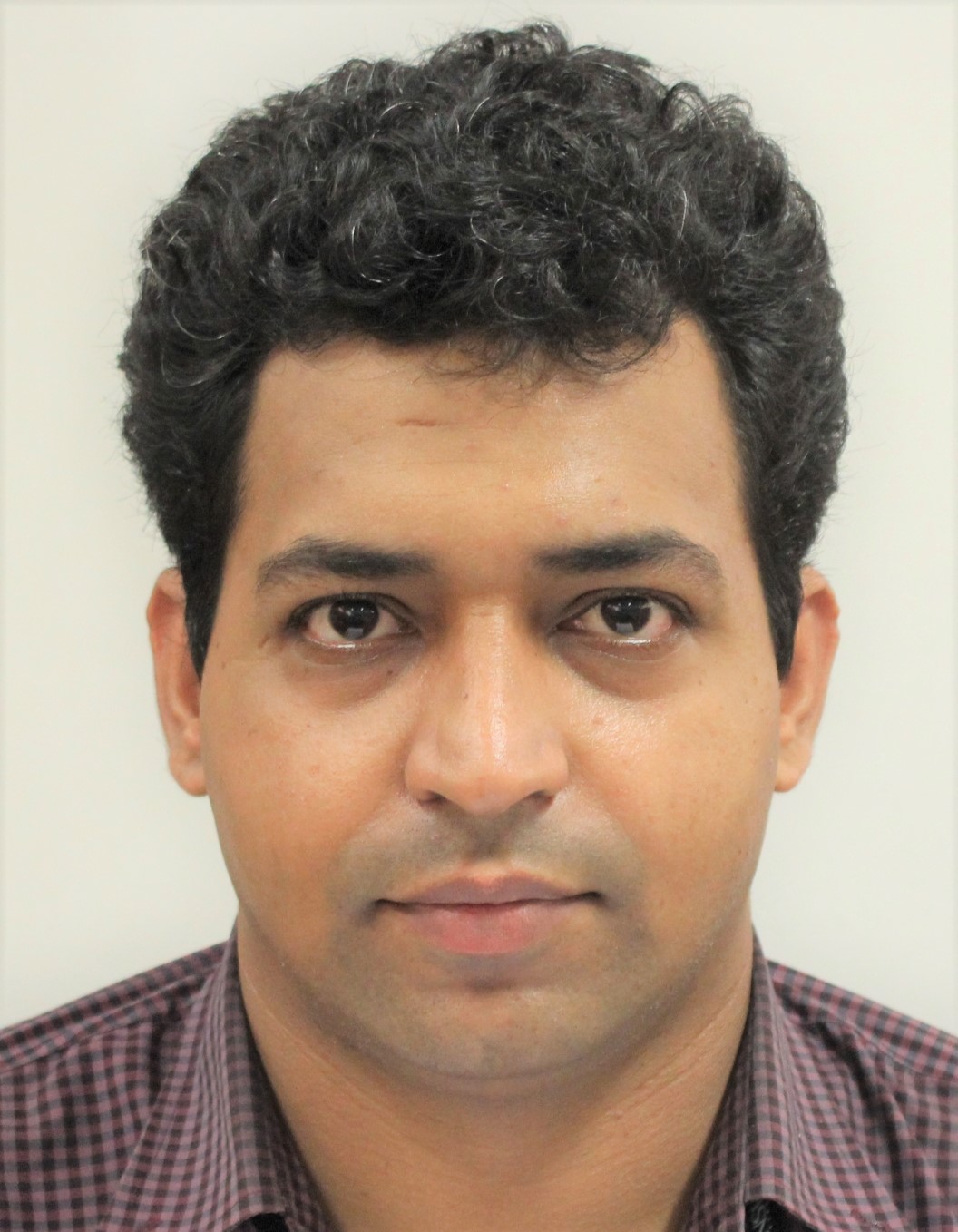}}]{Hafiz Suliman Munawar} is a Ph.D. student at the University of New South Wales (UNSW), Australia. He is a
multi-disciplinary researcher with experience in machine learning, disaster management and artificial intelligence. Hafiz has several international publications in various journals and conferences and has actively been working on disaster management using machine learning. He has over 5 years of IT industrial experience as well.
\end{IEEEbiography}

\vskip -2\baselineskip plus -1fil
\begin{IEEEbiography}[{\vspace{1 mm} \includegraphics[width=1.5in,
height=1.3in,clip, keepaspectratio]{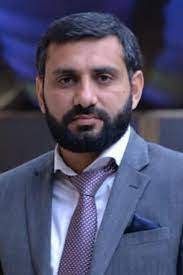}}]{Nasir Saeed} received his Bachelors of Telecommunication degree from University of Engineering and Technology, Peshawar, Pakistan, in 2009 and received Masters degree in satellite navigation from Polito di Torino, Italy, in 2012. He received his Ph.D. degree in electronics and communication engineering from Hanyang University, Seoul, South Korea in 2015. He was an Assistant Professor at the Department of Electrical Engineering, Gandhara Institute of Science and IT, Peshawar, Pakistan from August 2015 to September 2016. Dr. Saeed worked as an assistant professor at IQRA National University, Peshawar, Pakistan from October 2017 to July 2017. He is currently a Postdoctoral Research Fellow in King Abdullah University of Science and Technology (KAUST). His current areas of interest include cognitive radio networks, underwater optical wireless communications, dimensionality reduction, and localization.
\end{IEEEbiography}

\vskip -2\baselineskip plus -1fil
\begin{IEEEbiography}[{\vspace{1 mm} \includegraphics[width=1in,
height=1.3in,clip, keepaspectratio]{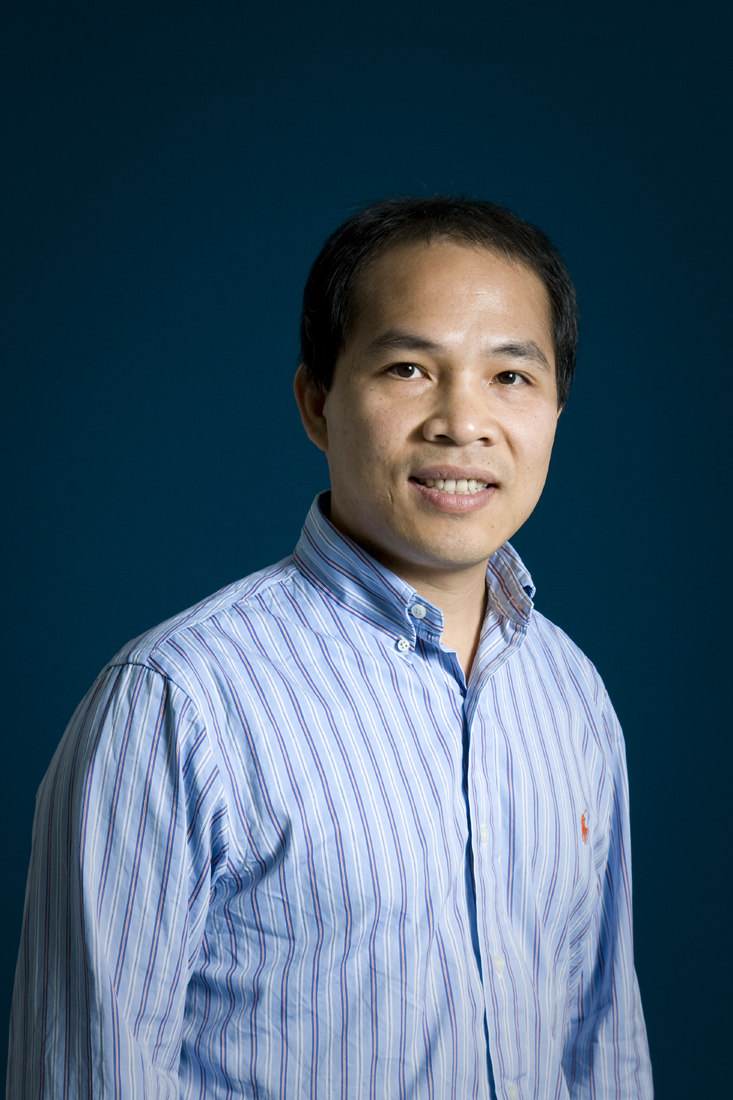}}]{Khoa N. Le} received his Ph.D. in October 2002 from Monash University, Melbourne, Australia. From April 2003 to June 2009, He was a Lecturer at Griffith University, Gold Coast campus, Griffith School of Engineering. From January to July 2008, he was a visiting professor at Intelligence Signal Processing Laboratory, Korea University, Seoul, Korea. From January 2009 to February 2009, he was a visiting professor at the Wireless Communication Centre, University Technology Malaysia, Johor Bahru, Malaysia. He is currently Associate Professor at School of Engineering, Design and Built Environment, Kingswood, Western Sydney University. His research interests are in wireless communications theory with applications to structural, construction management problems, image processing and wavelet theory. Dr. Le has been Editor for IEEE Transactions on Vehicular Technology, IEEE Wireless Communication Magazine, and IET Signal Processing. He is Bayu Chair Professor, Chongqing University of Science and Technology, Chongqing, China, 2020-2022. 
\end{IEEEbiography}

\end{document}